\documentclass[%
 reprint,
superscriptaddress,
 amsmath,amssymb,
 aps, longbibliography
]{revtex4-1}

\usepackage{graphicx}
\usepackage{dcolumn}
\usepackage{bm}
\usepackage{mathtools}
\usepackage{makecell}
\usepackage[colorinlistoftodos]{todonotes}
\usepackage{cancel}
\usepackage{comment}

\usepackage{amsmath}

\begin{document}


\title{Counterpropagating continuous variable entangled states in lossy coupled-cavity optical waveguides   }

\author{Hossein Seifoory}
\email{hossein.seifoory@queensu.ca}
\author{L. G. Helt}
\affiliation{Department of Physics, Engineering Physics and Astronomy, Queen's University, Kingston, Ontario K7L 3N6, Canada}
\author{J. E. Sipe}%
\affiliation{Department of Physics and Institute for Optical Sciences, University of Toronto, 60 St. George Street, Toronto, Ontario M5S 1A7, Canada}%
\author{Marc M. Dignam}
\affiliation{Department of Physics, Engineering Physics and Astronomy, Queen's University, Kingston, Ontario K7L 3N6, Canada}

\begin{abstract}
We present an integrated source of counterpropagating entangled states based on a coupled resonator optical waveguide that is pumped by a classical pulsed source incident from above the waveguide. We investigate theoretically the generation and propagation of continuous variable entangled states in this coupled-cavity system in the presence of intrinsic loss. Using a tight-binding approximation, we derive analytic time-dependent expressions for the number of photons in each cavity, as well as for the correlation variance between the photons in different pairs of cavities, to evaluate the degree of quantum entanglement. We also derive simple approximate expressions for these quantities that can be used to guide the design of such systems, and discuss how pumping configurations and physical properties of the system affect the photon statistics and the degree of quantum correlation.    
\end{abstract}

\maketitle


\section{\label{sec:level0}Introduction}
 Entangled quantum states have potential applications in quantum teleportation~\cite{1997/12/11/online,PhysRevLett.88.017903}, quantum computation, and quantum information~\cite{cerf2007quantum,bouwmeester2013physics}. They can either involve discrete variables (DVs), such as the polarization of a photon, or continuous variables (CVs), such as the quadratures of a beam of light.  Although DV systems provide high-fidelity operations, photonic-based DV entanglement is currently limited by the difficulties of single-photon generation and detection, and by high sensitivity to optical losses. In contrast, CV entanglement is more robust to loss, and can be more efficiently created and used for the implementation of quantum protocols~\cite{Huang2016,PhysRevA.83.042312,PhysRevX.5.041010,PhysRevLett.93.250503,PhysRevA.80.050303}. Spontaneous parametric down conversion (SPDC), a second order nonlinear process in which a pump photon is converted into a signal and an idler photon, is one of the processes that can be used to generate quantum correlated states~\cite{PhysRevLett.97.223602,PhysRevA.74.013815,Yang:07}. It has been implemented in both bulk media and integrated photonic structures. However, as the size and complexity of quantum information processing systems increase, the limitations in achieving stability, precision, and small physical size with bulk optical systems become significant. Systems for on-chip SPDC,  which are integrable with other photonic elements and could be used to generate CV entangled states involving two spatially separated sites, are therefore very promising~\cite{PhysRevA.83.062310,PhysRevA.82.012105,PhysRevA.83.032102}. 

One such platform involves the use of waveguides made of materials with a large second order nonlinear optical response, such as AlGaAs, to generate counterpropagating, quantum correlated photons~\cite{doi:10.1080/09500340802192431,Orieux:11}. The particular system we consider here is the coupled-resonator optical waveguide (CROW), in which the waveguide consists of optical cavities weakly coupled in one dimension. By adjusting the nature of the cavities and the coupling between them, the dispersive properties of the propagating modes can be controlled~\cite{Yariv:99}. Loss, which can destroy the nonclassical properties of light~\cite{PhysRevLett.55.2409,Jasperse:11,Seifoory:17,PhysRevA.97.023840,PhysRevA.85.052330}, can also be controlled to some extent, allowing at least a partial optimization for particular applications. CROW structures have been shown to have potential in generating CV entangled states between two side cavities coupled to the CROW, and as well between spatially separated sites~\cite{PhysRevA.83.062310}. It is the latter application we study here.

\begin{figure}
\includegraphics[width=\linewidth]{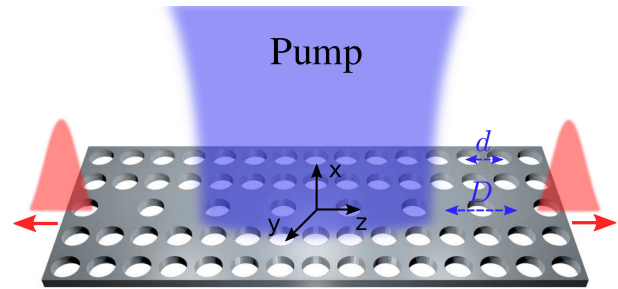}
\caption{Schematic picture of the particular CROW structure with period $D$ formed from defects in a slab photonic crystal with a square lattice of period $d$ and height $h$. The blue region shows the region covered by the pump. The origin of the coordinate system is at the center of the slab; i.e., the center of the central cavity.}
\label{fig:schematic}
\end{figure}

Our integrated source of entangled states is schematically shown in Fig.~\ref{fig:schematic}. A pump pulse is incident on a set of central cavities from above. Consequently, in order for the phase matching condition to be fulfilled, the generated signal and idler modes propagate in opposite directions in the CROW structure. An important advantage of such a configuration is the absence of the pump mode in the guided direction. Moreover, it has also been shown that the properties of the counterpropagating guided signal and idler modes can be tuned using the spectral and spatial properties of the pump~\cite{Orieux:11,PhysRevA.67.053810,PhysRevA.70.052317}.\par
The tight-binding~(TB) method~\cite{ashcroft2011solid,doi:10.1063/1.2737430}, which uses localized single-cavity modes as a basis, can be applied to model the evolution of light in such a coupled structure. Assuming that all the cavities are identical and support the same mode with complex frequency $\tilde{\omega}_F$, it has been shown~\cite{complex}  that in the nearest-neighbor tight-binding~(NNTB) approximation the dispersion relation can be written as
\begin{align}
\tilde{\omega}_{Fk}&\approx\tilde{\omega}_F[1-\tilde{\beta}_1 \cos(kD)]\nonumber\\
&\equiv\omega_{Fk}-i\gamma_{Fk},
\label{eq:dispersion_NNTB}
\end{align}
where $ \tilde{\beta}_1 $, $D$, and $k$ are respectively the complex coupling parameter, the periodicity of the CROW, and the Bloch vector component.  The imaginary part of the complex frequency is associated with the loss of the Bloch modes in the CROW. It is clear from Eq.~(\ref{eq:dispersion_NNTB}) that these modes experience different loss rates; it has been shown that the rates can differ by an order of magnitude or more~\cite{MohsenThesis,Fussell:07,PhysRevA.97.023840}.

In our previous work, we focused on the time evolution of a  state generated in a coupled-cavity system, and studied the evolution and propagation of squeezing and entanglement~\cite{PhysRevA.97.023840}. We presented analytic expressions for a general initial state, but only presented detailed results for an initial state that was a squeezed vacuum state in one of the cavities. In this work, we investigate both the generation and propagation of entangled states in coupled-cavity systems. In addition, we engineer the pump parameters to produce counterpropagating pulses of the generated signal and idler modes, which are entangled but are not individually squeezed. Including the effects of intrinsic propagation loss, we calculate the number of photons in each cavity and the CV correlation variance of photons in different cavities.

Previous approaches for generating counterpropagating entangled states have focused on photon pairs and been based either on ridge waveguides with vertical pumping~\cite{doi:10.1080/09500340802192431,Orieux:11}, or on periodic waveguides with horizontal pumping~\cite{PhysRevLett.118.183603}. The new approach of using a CROW has a number of advantages. First, the CROW allows us to control the group velocity and the frequency at which there is zero group velocity dispersion. Second, because a CROW can be modelled using a TB method, we are able to specify and model the effects of intrinsic scattering loss on the generated CV entanglement as a function of propagation distance, which is important for any application. Finally, using vertical pumping leads to counterpropagating entangled states, with no co-propagating pump at the outputs.

This paper is organized as follows. In Sec.~\ref{sec:level1} we present the general theory of the generation and the evolution of the generalized two mode squeezed state in lossy CROWs via SPDC. In Sec.~\ref{sec:result_for_gaussian}, we consider the special case of a pump that is Gaussian in time and space, and derive analytic expressions for the time dependence of the number of photons and the CV correlations. In Sec.~\ref{sec:level4}, we present our results for a particular CROW in a slab photonic crystal, and discuss how they might be affected by the pumping configuration and the physical properties of the structure. Finally, in Sec.~\ref{sec:conclusion}, we present our conclusions.

\section{\label{sec:level1}General Theory}
In order to determine both the generation and evolution of the entangled squeezed states in the system, we divide the analysis into two separate tasks. First, we study the creation of the entangled photons via SPDC using the backward Heisenberg method~\cite{PhysRevA.77.033808}, which is intrinsically a lossless approach. Having determined the initial entangled states created by the pump, we then include the loss to see how the generated state evolves in time and how loss affects it. Note that this two-step approach is valid because, for the parameters considered in this paper, the pump pulse is short enough that the signal loss is negligible over its duration.

\subsection{Generation}
There are two mode types that are relevant here, the fundamental modes and the pump modes; we indicate them by $ F $ and $ S $ respectively. In the SPDC process, two photons are generated in $ F $ modes from one pump photon in mode $ S $. Expanding the full displacement field $ \bm{D}(\bm{r}) $ in terms of the modes of interest, we have 
\begin{align}
\mathbf{D}\left(\mathbf{r}\right)&=\left(\int\text{d}k\,\sqrt{\frac{\hbar\omega_{Fk}}{2}}\mathbf{D}_{Fk}\left(\mathbf{r}\right)\hat{a}_{Fk}\right.\nonumber\\
&\quad\left.+\sum_{m}\int\text{d}\mathbf{q}\,\sqrt{\frac{\hbar\omega_{Sm\mathbf{q}}}{2}}\mathbf{M}_{Sm\mathbf{q}}\left(\mathbf{r}\right)\hat{a}_{Sm\mathbf{q}}\right)+\text{H.c.},
\label{eq:D_field_tot}          
\end{align} 
where $\mathbf{M}_{Sm\mathbf{q}}$, $\omega_{Sm\mathbf{q}}$, and $\hat{a}_{Sm\mathbf{q}}$ are the modes, eigenfrequencies and annihilation operators of the pump field, respectively, and $\mathbf{D}_{Fk}$, $\omega_{Fk}$, and $\hat{a}_{Fk}$ are the corresponding quantities for the generated signal and idler fields. Note that the integral over $k$ in Eq.~(\ref{eq:D_field_tot}) and in the rest of the paper (except where explicitly noted) only ranges from $-\pi/D$ to $\pi/D$. The continuous index, $\mathbf{q}$, is to identify the different pump modes in 3D while $m$ identifies the polarization state. The normalization conditions for the modes are presented in Appendix~\ref{appendix_normalization}. Here
\begin{align}
\left[\hat{a}_{Fk},\hat{a}^\dagger_{Sm\mathbf{q}}\right]&=\left[\hat{a}_{Fk},\hat{a}_{Sm\mathbf{q}}\right]=0,\nonumber\\ \left[\hat{a}_{Sm\mathbf{q}},\hat{a}^\dagger_{Sm\mathbf{q^\prime}}\right]&=\delta_{m\,m^\prime}\delta(\mathbf{q}-\mathbf{q^\prime}),\nonumber\\
\left[\hat{a}_{Fk},\hat{a}^\dagger_{Fk^\prime}\right]&=\delta(k-k^\prime).
\end{align}
For convenience we put
\begin{equation}
\hat{a}_{Fk}    \rightarrow \hat{b}_{k},\quad \hat{a}_{Sm\mathbf{q}}        \rightarrow \hat{c}_{m\mathbf{q}},
\end{equation}
and the linear Hamiltonian is then given by
\begin{equation}
H_{L}=\int\text{d}k\,\hbar\omega_{Fk}\hat{b}_{k}^{\dagger}\hat{b}_{k}+\sum_{m}\int\text{d}\mathbf{q}\,\hbar\omega_{Sm\mathbf{q}}\hat{c}_{m\mathbf{q}}^{\dagger}\hat{c}_{m\mathbf{q}},
\end{equation}
where we neglect the zero point energy and use only the real part of $\tilde{\omega}_{Fk}$ for the mode frequency. The nonlinear Hamiltonian that should be added to $H_{L}$ to construct the full Hamiltonian is~\cite{PhysRevA.77.033808}
\begin{equation}
H_{NL}=-\sum_{m}\int\text{d}k_{1}\text{d}k_{2}\text{d}\mathbf{q}\,S\left(k_{1},k_{2},m,\mathbf{q}\right)\hat{b}_{k_{1}}^{\dagger}\hat{b}_{k_{2}}^{\dagger}\hat{c}_{m\mathbf{q}}+\text{H.c.},
\label{H_NL}
\end{equation}
where $ S\left(k_{1},k_{2},m, \mathbf{q}\right) $ is the coupling coefficient, which is given by
\begin{align}
\label{eq:S_function01}
S\left(k_{1},k_{2},m, \mathbf{q}\right) &=\frac{1}{\varepsilon_{0}}\sqrt{\frac{\hbar\omega_{Fk_1}\hbar\omega_{Fk_2}\hbar\omega_{Sm\mathbf{q}}}{8}}\int\text{d}\mathbf{r}\,\chi_{2}^{ijk}\left(\mathbf{r}\right)\nonumber\\
&\quad\times\frac{\left[D_{Fk_{1}}^{i}\left(\mathbf{r}\right)D_{Fk_{2}}^{j}\left(\mathbf{r}\right)\right]^{*}M^k_{Sm\mathbf{q}}\left(\mathbf{r}\right)}{\varepsilon_{0}n^{2}\left(\mathbf{r};\omega_{Fk_1}\right)n^{2}\left(\mathbf{r};\omega_{Fk_2}\right)n^{2}\left(\mathbf{r};\omega_{Sm\mathbf{q}}\right)},
\end{align}
where $n\left(\mathbf{r};\omega\right)$ is a real, position and frequency dependent refractive index and $\chi_{2}^{ijk}\left(\mathbf{r}\right)$ is the position-dependent second-order nonlinear susceptibility.\par
We take the pump to be a classical pulse incident on the slab, which we can expand as a superposition of the pump modes $\bm{M}_{Sm\mathbf{q}}\left(\bm{r}\right)$. We borrow a strategy from Yang \textit{et al}.~\cite{PhysRevA.77.033808} and define asymptotic-in and -out states to be respectively the input and output states of the nonlinear region at $t=0$, taking $t=0$ to be the time when the pump is centred on the slab. For the asymptotic-in state $\left|\psi_{\text{in}}\right\rangle$ that describes the classical pump pulse as a coherent state, we have
\begin{equation}
\left|\psi_{\text{in}}\right\rangle =e^{\alpha\sum_{m}\int\text{d}\mathbf{q}\,\phi_{P}\left(m,\mathbf{q}\right)\hat{c}^\dag_{m\mathbf{q}}-\text{H.c.}}\left|\text{vac}\right\rangle ,
\label{eq:psi_in}
\end{equation} 
where $\alpha$ is a complex number, and we normalize the complex function $\phi_{P}\left(m,\mathbf{q}\right)$ according to 
\begin{equation}
\label{eq:normalization_condition}
    \sum_{m}\int \text{d}\mathbf{q}\,|\phi_P\left(m,\mathbf{q}\right)|^{2}=1.
\end{equation}
The expectation value of the displacement field of the pump pulse is then 
\begin{align}
\langle  \psi_{\text{in}}|\mathbf{D}\left( \mathbf{r}\right)|\psi_{\text{in}}\rangle=&\alpha\sum_{m}\int\text{d}\mathbf{q}\,\sqrt{\frac{\hbar\omega_{Sm\mathbf{q}}}{2}}\phi_{P}\left(m,\mathbf{q}\right)\mathbf{M}_{Sm\mathbf{q}}\left(\mathbf{r}\right)\nonumber\\
&+\text{c.c.}
\end{align}
and since 
\begin{equation}
    \langle  \psi_{\text{in}}|\hat{c}^\dag_{m\mathbf{q}}\hat{c}_{m\mathbf{q}}|\psi_{\text{in}}\rangle=|\alpha|^2|\phi_P(m,\mathbf{q})|^2
\end{equation}
we can identify $|\alpha|^2$ as the expectation value of the number of photons in the pump pulse. Following the backward Heisenberg picture approach~\cite{PhysRevA.77.033808}, the asymptotic-out state for the \textit{generated} photons in the first approximation is then
\begin{equation}
\label{assym_out}
\left|\psi_{\text{out}}^{F}\right\rangle =e^{\frac{\beta}{\sqrt{2}}\int\text{d}k_{1}\text{d}k_{2}\,\phi\left(k_{1},k_{2}\right)\hat{b}_{k_{1}}^{\dagger}\hat{b}_{k_{2}}^{\dagger}-\text{H.c.}}\left|\text{vac}\right\rangle ,
\end{equation}
where $\phi\left(k_{1},k_{2}\right)$ is the biphoton wave function, which from Yang \textit{et al}.~\cite{PhysRevA.77.033808} is given by
\begin{align}
\label{eq:biphoton_sipe}
\phi\left(k_{1},k_{2}\right)&=\frac{2i\sqrt{2}\pi\alpha}{\beta\hbar}\sum_{m}\int\text{d}\mathbf{q}\,\phi_{P}\left(m,\mathbf{q}\right)\nonumber\\
&\quad\times S\left(k_{1},k_{2},m, \mathbf{q}\right) \delta (\omega_{Sm\mathbf{q}}-\omega_{F_{k_{1}}}-\omega_{F_{k_{2}}}),
\end{align}
where $\beta$ is a real and positive normalization constant chosen to ensure that $\int \text{d}k_1\,\text{d}k_2\,|\phi\left(k_{1},k_{2}\right)|^{2}=1$.

While the biphoton wave function in general obeys the symmetry $\phi\left(k_{1},k_{2}\right)=\phi\left(k_{2},k_{1}\right)$, it can sometimes be useful to work with a function that breaks this symmetry to focus attention on a particular quadrant of $\left(k_{1},k_{2}\right)$ space. In this work, we will always choose the pump parameters such that to a very good approximation, $\phi(k_1,k_2)$ is nonzero only when $k_1$ and $k_2$ have opposite signs, as we detail in Section~\ref{sec:result_for_gaussian} below. Then, we can write 
\begin{align}
&\int_{-\pi/D}^{\pi/D}\text{d}k_{1}\int_{-\pi/D}^{\pi/D}\text{d}k_{2}\,\phi\left(k_{1},k_{2}\right)\hat{b}_{k_{1}}^{\dagger}\hat{b}_{k_{2}}^{\dagger}\nonumber\\
&\approx\int_{0}^{\pi/D}\text{d}k_{1}\int_{-\pi/D}^{0}\text{d}k_{2}\,\phi\left(k_{1},k_{2}\right)\hat{b}_{k_{1}}^{\dagger}\hat{b}_{k_{2}}^{\dagger}\nonumber\\
&\quad+\int_{-\pi/D}^{0}\text{d}k_{1}\int_{0}^{\pi/D}\text{d}k_{2}\,\phi\left(k_{1},k_{2}\right)\hat{b}_{k_{1}}^{\dagger}\hat{b}_{k_{2}}^{\dagger}\nonumber\\
&=2\int_{0}^{\pi/D}\text{d}k_{1}\int_{-\pi/D}^{0}\text{d}k_{2}\,\phi\left(k_{1},k_{2}\right)\hat{b}_{k_{1}}^{\dagger}\hat{b}_{k_{2}}^{\dagger}.
\end{align}
We then define
\begin{equation}
\label{eq:Luke_Phi}
\Phi\left(k_{1},k_{2}\right)\equiv\sqrt{2}\phi\left(k_{1},k_{2}\right)\Theta\left(k_{1}\right)\Theta\left(-k_{2}\right),
\end{equation}
where $ \Theta\left(k\right) $ is the Heaviside function, such that we may rewrite Eq.~\eqref{assym_out} as
\begin{equation}
\label{eq:assym_out_2}
\left|\psi_{\text{out}}^{F}\right\rangle =e^{\beta\int\text{d}k_{1}\text{d}k_{2}\,\Phi\left(k_{1},k_{2}\right)\hat{b}_{k_{1}}^{\dagger}\hat{b}_{k_{2}}^{\dagger}-\text{H.c.}}\left|\text{vac}\right\rangle.
\end{equation}

 Employing a Schmidt decomposition~\cite{peres2006quantum,doi:10.1119/1.17904}, we have
 \begin{equation}\label{eq:phi_inermsof_mu_nu}
\Phi\left(k_{1},k_{2}\right)=\sum_{\lambda}\sqrt{p_{\lambda}}\mu_{\lambda}\left(k_{1}\right)\nu_{\lambda}\left(k_{2}\right),
 \end{equation}
for $p_{\lambda} > 0$ with $\sum_{\lambda}p_{\lambda}=1$, where the Schmidt functions are orthonormal, 
\begin{equation}
\int\text{d}k\mu_{\lambda}(k)\mu^*_{\lambda^\prime}(k)=\int\text{d}k\nu_{\lambda}(k)\nu^*_{\lambda^\prime}(k)=\delta_{\lambda,\lambda^{\prime}}.
\end{equation}
We extend the sets of ${\mu_\lambda(k)}$ and ${\nu_\lambda(k)}$ associated with $p_\lambda>0$ to form complete sets with 
\begin{equation}
    \label{eq:orthonormal}
    \sum_\lambda \mu_\lambda(k)\mu^*_\lambda(k^\prime)=\sum_\lambda \nu_\lambda(k)\nu^*_\lambda(k^\prime)=\delta(k-k^\prime),
\end{equation}
and with some of the $p_\lambda$ appearing in Eq.~(\ref{eq:phi_inermsof_mu_nu}) then equal to zero. Using Eqs.~\eqref{eq:assym_out_2} and \eqref{eq:phi_inermsof_mu_nu}, the generated squeezed state can be written as
\begin{equation}
\left|\psi_{\text{out}}^{F}\right\rangle =\hat{S}\left|\text{vac}\right\rangle ,
\end{equation}  
where the squeezing operator, $ \hat{S} $, can be written as
\begin{align}
\hat{S}&=\exp\left(\beta\int\text{d}k_{1}\text{d}k_{2}\,\sum_{\lambda}\sqrt{p_{\lambda}}\mu_{\lambda}\left(k_{1}\right)\nu_{\lambda}\left(k_{2}\right)\right.\nonumber\\
&\quad\times\left.\hat{b}_{k_{1}}^{\dagger}\hat{b}_{k_{2}}^{\dagger}-\text{H.c.}\vphantom{\beta\int\text{d}k_{1}\text{d}k_{2}\,\sum_{\lambda}}\right)\nonumber\\
&=\exp\left(\sum_{\lambda}r_{\lambda}\hat{B}_{\lambda}^{\dagger}\hat{C}_{\lambda}^{\dagger}-\sum_{\lambda}r_{\lambda}^{*}\hat{B}_{\lambda}\hat{C}_{\lambda}\right),
\end{align}
where $r_{\lambda}=\beta\sqrt{p_{\lambda}}$ is the squeezing parameter,
\begin{align}
\label{eq:B_operator}
\hat{B}_{\lambda}&\equiv\int\mu_{\lambda}^{*}\left(k\right)\hat{b}_{k}\text{d}k,
\end{align}
and
\begin{align}
\label{eq:C_operator}
\hat{C}_{\lambda}&\equiv\int\nu_{\lambda}^{*}\left(k\right)\hat{b}_{k}\text{d}k.
\end{align}
Using Eq.~(\ref{eq:orthonormal}), it can be shown that $ [\hat{B}_{\lambda},\hat{B}_{\lambda^{\prime}}^{\dagger}]=[\hat{C}_{\lambda},\hat{C}_{\lambda^{\prime}}^{\dagger}]=\delta_{\lambda,\lambda^{\prime}} $ and $ [\hat{B}_{\lambda},\hat{C}_{\lambda^{\prime}}^{\dagger}]=[\hat{B}_{\lambda},\hat{B}_{\lambda^{\prime}}]=[\hat{C}_{\lambda},\hat{C}_{\lambda^{\prime}}]=[\hat{B}_{\lambda},\hat{C}_{\lambda^{\prime}}]=0$. \par
The importance of the Schmidt decomposition and the operator transformation is that it enables us to express the generated state as a generalized two-mode squeezed state, where the modes are no longer the Bloch modes. As we shall see in the next section, this will enable us to easily determine the evolution of the state in the presence of loss.
\subsection{Evolution}
As mentioned earlier, we have assumed that the loss during the generation process is negligible. However, the effect of loss cannot be ignored when calculating the evolution of the generated pulses down the CROW. \par
Following the formalism presented in our previous work~\cite{PhysRevA.97.023840} on lossy coupled-cavity systems, the individual single-mode cavity annihilation operator for the $ p^{th} $ cavity, $ \hat{a}_p $, can be written in terms of the $ k^{th} $ mode annihilation operator of the coupled-cavity-system, $ \hat{b}_k $, as
\begin{equation}
\label{a_to_b}
\hat{a}_{p}(t)=\sqrt{\frac{D}{2\pi}}\int\hat{b}_{k}(t)e^{ikpD}\text{d}k. 
\end{equation}
The time evolution of the full coupled-cavity annihilation operator can also be found by solving the adjoint master equation for this open, lossy system~\cite{open_quantum_system}. We have previously shown that the time dependence of the individual annihilation operators is given by
\begin{equation}
\label{b_time}
\hat{b}_k(t)=\hat{b}_ke^{-i\tilde{\omega}_{Fk}t},
\end{equation}
where $ \hat{b}_k=\hat{b}_k(0) $ is the corresponding operator in the Schr\"{o}dinger representation~\cite{PhysRevA.85.013809}.
Using Eqs.~(\ref{a_to_b}), (\ref{b_time}), and their complex conjugates, the time dependent average photon number in the $ p^{th} $ cavity can be written as 
\begin{align}
\label{eq:apdap}
\left\langle \hat{a}_{p}^{\dagger}\left(t\right)\hat{a}_{p}\left(t\right)\right\rangle &=\frac{D}{2\pi}\int\int \text{d}k\text{d}k^{\prime}\left\langle \hat{b}_{k}^{\dagger}\hat{b}_{k^{\prime}}\right\rangle e^{-i\left(k-k^{\prime}\right)pD}\nonumber\\
&\times\left(e^{i\tilde{\omega}_{F}^{*}\left(1-\tilde{\beta}_{1}^{*}\cos(kD)\right)t}e^{-i\tilde{\omega}_{F}\left(1-\tilde{\beta}_{1}\cos(k^{\prime}D)\right)t}\right),
\end{align} 
where we have used the lossy dispersion relation of the CROW structure~[Eq.~\eqref{eq:dispersion_NNTB}]. To facilitate the evaluation of $\left\langle \hat{b}_{k}^{\dagger}\hat{b}_{k^{\prime}}\right\rangle$, we introduce the restricted operators,
\begin{align}
\hat{b}_{k,+}&\equiv\Theta\left(k\right)\hat{b}_{k}\nonumber\\
\hat{b}_{k,-}&\equiv\Theta\left(-k\right)\hat{b}_{k}.
\end{align}
Using these operators, we can write 
\begin{equation}
    \label{eq:4_terms}
    \left\langle \hat{b}_{k}^{\dagger}\hat{b}_{k^{\prime}}\right\rangle=
    \left\langle \hat{b}_{k,+}^{\dagger}\hat{b}_{k^{\prime},-}^{}+
    \hat{b}_{k,-}^{\dagger}\hat{b}_{k^{\prime},-}^{}+
    \hat{b}_{k,+}^{\dagger}\hat{b}_{k^{\prime},+}^{}+
    \hat{b}_{k,-}^{\dagger}\hat{b}_{k^{\prime},+}^{}\right\rangle.
\end{equation}
To evaluate each of these terms, we use the following Bogoliubov transformations
\begin{equation}
\begin{multlined}
\label{bogol1}
\hat{S}^{\dagger}\hat{b}_{k,+}\hat{S}=\hat{S}^{\dagger}\sum_{\lambda}\mu_{\lambda}\left(k\right)\hat{B}_{\lambda}\hat{S}\\=\sum_{\lambda}\mu_{\lambda}\left(k\right)\left[\hat{B}_{\lambda}\cosh\left(r_\lambda\right)-\hat{C}_{\lambda}^{\dagger}\sinh\left(r_\lambda\right)\right], 
\end{multlined}
\end{equation}
\begin{equation}
\begin{multlined}
\label{bogol2}
\hat{S}^{\dagger}\hat{b}_{k,-}\hat{S}=\hat{S}^{\dagger}\sum_{\lambda}\nu_{\lambda}\left(k\right)\hat{C}_{\lambda}\hat{S}\\=\sum_{\lambda}\nu_{\lambda}\left(k\right)\left[\hat{C}_{\lambda}\cosh\left(r_\lambda\right)-\hat{B}_{\lambda}^{\dagger}\sinh\left(r_\lambda\right)\right].
\end{multlined}
\end{equation}
Using these in Eq.~\eqref{eq:4_terms}, we obtain
\begin{align}
\label{eq:bdkbk}
\left\langle \hat{b}_{k}^{\dagger}\hat{b}_{k^{\prime}}\right\rangle &=\sum_{\lambda}\bigg(\mu_{\lambda}^{*}\left(k\right)\mu_{\lambda}\left(k^{\prime}\right)\nonumber+\nu_{\lambda}^{*}\left(k\right)\nu_{\lambda}\left(k^{\prime}\right)\bigg)\\
&\times\sinh^{2}\left(r_\lambda\right).
\end{align} 
To study the degree of entanglement between the photons in cavities $ p $ and $ p^\prime $ in a CROW, we use the correlation variance, which is defined as
\begin{equation}
\Delta_{p,p^\prime}^2=\left\langle [\Delta(\hat{X}_p-\hat{X}_{p^\prime})]^2\right\rangle+ \left\langle [\Delta(\hat{Y}_p+\hat{Y}_{p^\prime})]^2\right\rangle,
\label{eq:correlation_nm_def}
\end{equation}  
where 
\begin{equation}
\label{X_and_Y}
\begin{aligned}
\hat{X}_p&\equiv\hat{a}_p+\hat{a}_p^{\dagger},\\
\hat{Y}_p&\equiv-i(\hat{a}_p-\hat{a}_p^{\dagger}). 
\end{aligned}
\end{equation}
It has been shown that $ \Delta_{p,p^\prime}^2  <4$ can be considered as the inseparability criterion for entanglement~\cite{PhysRevLett.84.2722,PhysRevLett.84.2726,Masada2015,Zhang2015}. Using Eq.~(\ref{X_and_Y}) in Eq.~(\ref{eq:correlation_nm_def}), the time-dependent correlation variance can be written as
\begin{equation}
\label{eq:entanglement_01}
\Delta_{pp^\prime}^{2}=4+4\left(\langle \hat{a}_{p}^{\dagger}\hat{a}_{p}\rangle +\langle \hat{a}_{p^{\prime}}^{\dagger}\hat{a}_{p^{\prime}}\rangle -\langle \hat{a}_{p}\hat{a}_{p^{\prime}}\rangle-\langle \hat{a}_{p}^{\dagger}\hat{a}_{p^{\prime}}^{\dagger}\rangle \right).
\end{equation}\par
Following a procedure similar to that used to arrive at Eqs.~(\ref{eq:apdap}) and (\ref{eq:bdkbk}), one can derive the other expectation values that are needed to evaluate the variances of the quadrature operators and the correlation variance in the CROW structure.\par 
\section{Results for a Gaussian pump pulse} \label{sec:result_for_gaussian}
The results of the previous sections are general and independent of the temporal and spatial form of the pump pulse, as long as $\phi(k_1,k_2)$ is nonzero only when $k_1$ and $k_2$ have opposite signs. However, in this section we consider the special case of a Gaussian pump pulse incident on the slab, and brought to a Gaussian focus there. \par
We assume that the slab does not have a significant effect on the pump pulse, and so take the pump modes to be plane waves in free space and set $n(\mathbf{r},\omega_{Sm\mathbf{q}})=1$ in Eq.~(\ref{eq:S_function01}). Thus, we have
\begin{equation}
\mathbf{M}_{Sm\mathbf{q}}\left(\mathbf{r}\right)=\frac{\sqrt{\varepsilon_{0}}\mathbf{e}_{m,\mathbf{q}}}{\left(2\pi\right)^{3/2}}e^{i\mathbf{q}\cdot\mathbf{r}},
\label{M}          
\end{equation}
where $\mathbf{e}_{m,\mathbf{q}}$ is the polarization unit vector. In what follows, we assume that in the vicinity of the CROW, the pump is polarized in the $y$-direction. Thus, we obtain
\begin{equation}
\phi_{P}\left(m,\mathbf{q}\right)=\delta_{my}\varphi\left(q_{x}\right)F\left(q_{y},q_{z}\right),
\end{equation}
where
\begin{equation}
F\left(q_{y},q_{z}\right)=\frac{1}{2\pi}\int\text{d}y\text{d}z\,f\left(y,z\right)e^{-i\left(q_{y}y+q_{z}z\right)},
\end{equation}
is the Fourier transform of the transverse profile
\begin{equation}
\label{eq:f_yz_text}
f\left(y,z\right)=\frac{\sqrt{2}}{\sqrt{\pi}W_{S}}e^{-\frac{y^{2}+z^{2}}{W_{S}^{2}}}e^{iq_{P}\frac{y^{2}+z^{2}}{2R_{P}}},
\end{equation}
and $q_P$ is the value of $q_x$ at which $\phi(q_x)$ peaks. Here $R_{P}$ and $W_{S}$ are the radius of curvature and spot size, respectively, evaluated at $x=0$ and $q_{x}=q_{P}$. We have assumed the Rayleigh range to be much larger than the slab thickness, which justifies the neglect of the Gouy phase. The prefactors have been chosen so that the normalization condition Eq.~(\ref{eq:normalization_condition}) becomes
\begin{equation}
    \int\text{d}q_{x}\left|\varphi\left(q_{x}\right)\right|^{2}=1.
\end{equation}
Neglecting the dependence of the indices of refraction and the frequencies under the square root on $k_{1}$, $k_{2}$, and $\mathbf{q}$, based on the small frequency range of the input pump pulse and the limited range of the signal and idler photons in the CROW, we can rewrite the biphoton wave function of Eq.~\eqref{eq:Luke_Phi} as 
\begin{equation}
\begin{split}
\label{eq:bithoton_approx}
\Phi &\left(k_{1},k_{2}\right)=\frac{i\alpha\sqrt{\pi}}{\beta\hbar\sqrt{\varepsilon_0}}\sqrt{\left(\hbar\omega_{F}\right)^{2}\hbar\omega_{Sy}}\int\text{d}q_{x}\varphi(q_{x})\\
&\times\int\text{d}\mathbf{r}\,\chi_{2}^{ijy}\left(\mathbf{r}\right)\frac{\left[D_{Fk_{1}}^{i}\left(\mathbf{r}\right)D_{Fk_{2}}^{j}\left(\mathbf{r}\right)\right]^{*}f(y,z)e^{iq_{x}x}}{\varepsilon_{0}n^{4}\left(\mathbf{r};\omega_{F}\right)}\\
&\times \delta (cq_{x}-\omega_{F_{k_{1}}}-\omega_{F_{k_{2}}})\Theta\left(k_{1}\right)\Theta\left(-k_{2}\right).
\end{split}
\end{equation}
Because we have made the approximation that the transverse profile of the pump does not depend on frequency (for the frequencies of interest) in Eq.~\eqref{eq:bithoton_approx} we set $\omega_{Sy\mathbf{q}}=cq_{x}$ in the Dirac delta function.

We now employ the nearest-neighbour tight-binding approximation~\cite{doi:10.1063/1.2737430} and expand the $D_{Fk}^{i}\left(\mathbf{r}\right)$ modes in terms of the single-cavity quasimodes, $N_{Fp}^{i}\left(\mathbf{r}\right)$, as
\begin{equation}
D_{Fk}^{i}\left(\mathbf{r}\right)=\sqrt{\frac{D}{2\pi}}\sum_{p}N_{Fp}^{i}\left(\mathbf{r}\right)e^{ikpD},
\label{eq:D_in terms of N}
\end{equation}
which leads to the lossy frequency dispersion given in Eq.~(\ref{eq:dispersion_NNTB}). The single-cavity quasimodes and frequencies are calculated in the standard way using finite difference time domain calculations~\cite{doi:10.1063/1.2737430}. Assuming that the cavity modes are well-localized~\cite{well_localized}, we obtain 
\begin{equation}\label{eq:biphoton_sigma_and_integral}
\begin{multlined}
\Phi\left(k_{1},k_{2}\right)=\frac{i \alpha D}{2\beta\hbar\sqrt{ \varepsilon_0\pi}}\sqrt{\left(\hbar\omega_{F}\right)^{2}\hbar\omega_{S}}\int\text{d}q_{x}\varphi(q_{x})\\
\times \sum_p \int\text{d}\mathbf{r}\,\chi_{2}^{ijy}\left(\mathbf{r}\right)\frac{\left[N_{Fp}^{i}\left(\mathbf{r}\right)N_{Fp}^{j}\left(\mathbf{r}\right)\right]^{*}f(y,z)e^{iq_{x}x}}{\varepsilon_{0}n^{4}\left(\mathbf{r};\omega_{F}\right)}\\
\times e^{-i(k_1 + k_2 )pD} \delta (cq_{x}-\omega_{F_{k_{1}}}-\omega_{F_{k_{2}}})\Theta\left(k_{1}\right)\Theta\left(-k_{2}\right).
\end{multlined}
\end{equation}
We define
\begin{equation}
\label{PHI_qx_text}
\varphi(q)=\sqrt{W_T/\sqrt{2\pi}}\exp{\left[-\left(\frac{\left(q-q_{P}\right)W_T}{2}\right)^{2}\right]},
\end{equation}
and, because and spatial extent of the single-cavity quasimodes is small relative to that of the pump field, in the integral in Eq.~\eqref{eq:biphoton_sigma_and_integral} we replace $f\left(y,z\right)$ by 
\begin{equation}
f\left(y=0,z=pD\right)\approx\frac{\sqrt{2}}{\sqrt{\pi}W_{S}}e^{-\frac{p^{2}D^{2}}{W_{S}^{2}}},
\end{equation}
to obtain the approximate expression
\begin{equation}
\label{biphoton_01}
\begin{multlined}
\Phi(k_{1},k_{2})=\frac{i\alpha\bar{\chi}_{2}}{\beta}\sqrt{\frac{\hbar\omega_{F}^{2}\omega_{S}W_T}{\varepsilon_{0}\left(2\pi\right)^{3/2}}}e^{\frac{-(k_{1}+k_{2})^{2}W_S^{2}}{4}}\\
\times\int\text{d}q_{x}\,e^{-\frac{\left(q_{x}-q_{P}\right)^{2}W_T^{2}}{4}}\delta(\omega_{Sq_{x}}-\omega_{Fk_{1}}-\omega_{Fk_{2}})\Theta\left(k_{1}\right)\Theta\left(-k_{2}\right),
\end{multlined}
\end{equation}
where 
\begin{equation}\label{eq:Xi2bar}
\bar {\chi}_{2}\equiv \int\text{d}\mathbf{r}\,\chi_{2}^{ijy}\left(\mathbf{r}\right)\frac{\left[N_{F0}^{i}\left(\mathbf{r}\right)N_{F0}^{j}\left(\mathbf{r}\right)\right]^{*}}{\varepsilon_{0}n^{4}\left(\mathbf{r};\omega_{F}\right)}e^{iq_{P}x},
\end{equation}
is the effective second order susceptibility for the system~\cite{expiq_x}~(See Appendix~\ref{appendix_B} for more details).\par
For $\omega_{Fk}\equiv\omega_{F}\left[1-\beta_{1}\cos\left(kD\right)\right]$, which we consider to be a real quantity at this point, Eq.~(\ref{biphoton_01}) can be rewritten as
\begin{widetext}
\begin{equation}
\label{biphoton_02}
\begin{multlined}
{\Phi}(k_{1},k_{2})=Q_0\int\text{d}q_{x}\,e^{\frac{-(k_{1}+k_{2})^{2}W_S^{2}}{4}}e^{-\frac{\left(q_{x}-q_{P}\right)^{2}W_T^{2}}{4}}\delta\left\{ q_{x}-\frac{1}{c}\left[\omega_{Fk_1}+\omega_{Fk_2}\right]\right\}\Theta\left(k_{1}\right)\Theta\left(-k_{2}\right)\\
=Q_{0}\exp\left(\frac{-(k_{1}+k_{2})^{2}W_S^{2}}{4}\right)\exp\left(-\left(\frac{2\omega_{F}-\beta_{1}\omega_{F}\left[\cos\left(k_{1}D\right)+\cos\left(k_{2}D\right)\right]-\omega_{P}}{2c}\right)^{2}W_T^{2}\right)\Theta\left(k_{1}\right)\Theta\left(-k_{2}\right),
\end{multlined}
\end{equation}
\end{widetext}
where
\begin{equation}\label{eq:Q_0}
Q_{0}\equiv\frac{i\alpha\bar{\chi}_{2}}{\beta c}\sqrt{\frac{\hbar\omega_{F}^{2}\omega_{S}W_T}{\varepsilon_{0}\left(2\pi\right)^{3/2}}}.
\end{equation}\par
\begin{figure*}
\begin{center}
\includegraphics[scale=0.38]{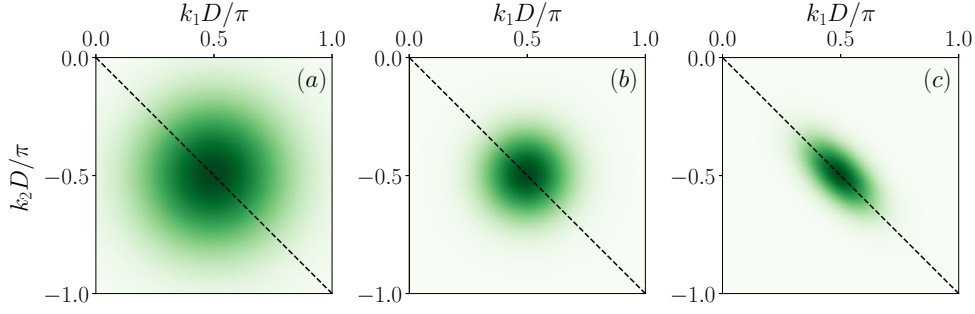}
\caption{Biphoton wave function of Eq.~\eqref{biphoton_02} for three different pumping configurations: (a)~ $\sigma_+\,D=\sigma_-\,D=0.47$, (b)~ $\sigma_+\,D=\sigma_-\,D=0.28$, and (c)~  $\sigma_+\,D=0.14$, $\sigma_-\,D=0.28$. In all these three cases we consider $k_0=\pi/2D$.} \label{fig:biphoton}
\end{center}
\end{figure*}
 \par 

In order to derive analytic expressions for the photon number and correlation variance as a function of cavity index, $p$, we need to place further restrictions on the pump pulse. From the first exponential in Eq.~(\ref{biphoton_02}), we see that the biphoton wave function will only be non-negligible if $k_2$ is approximately equal to $-k_1$. Thus we set 
\begin{equation}
\begin{aligned}
k_{1}   &\rightarrow k_{0}+\delta_{1},\\
k_{2}   &\rightarrow-k_{0}+\delta_{2}, 
\end{aligned}
\end{equation} 
where the $\delta_{i}$ are small relative to $W_S$, where $k_0$ is determined by the central frequency of the pump, through the equation 
\begin{equation}
\omega_{P}=2\omega_{Fk_{0}}=2\omega_{F}-2\beta_{1}\omega_{F}\cos\left(k_{0}D\right).
\end{equation}
In order to obtain a biphoton wave function for which there is an analytic Schmidt decomposition, we take $k_0=\pi/(2D)$ and choose the frequency width parameter, $W_T$, of the pulse to satisfy $c/W_T\ll\Delta$, where $\Delta=2\omega_{F}\beta_{1}$.
Now, expanding the cosines in Eq.~(\ref{biphoton_02}) to first order in $\delta_1$ and $\delta_2$, the biphoton wave function can be rewritten as 
\begin{align}
\label{biphoton_form}
 &{\Phi(k_0+\delta_1,-k_0+\delta_2)}=\nonumber\\
&\sqrt{\frac{2}{\pi\sigma_{+}\sigma_{-}}}
\exp\left(-\frac{(\delta_{1}+\delta_{2})^{2}}{2\sigma_{+}^{2}}\right)\exp\left(-\frac{\left(\delta_{1}-\delta_{2}\right)^{2}}{2\sigma_{-}^{2}}\right),
\end{align}
where
\begin{equation}
\sigma_{+}\equiv\frac{\sqrt{2}}{W_S}
\label{eq:sigma_plus}
\end{equation}
and
\begin{equation}
\label{eq:sigma_minus}
\sigma_{-}\equiv\frac{\sqrt{2}c}{W_T\beta_{1}\omega_{F}D\sin\left(|k_{0}|D\right)}.
\end{equation}
Strictly speaking, the biphoton wave function of Eq.~\eqref{biphoton_form} does not satisfy the restriction that it is zero unless $k_1>0$ and $k_2<0$.  However, as long as $\sigma_+$ and $\sigma_-$ are chosen to be small enough, then these conditions are satisfied to a very high degree.  In what follows, we shall only consider situations where this is the case. Using the normalization condition, $\int \text{d}k_{1}\,\text{d}k_{2}\left|{\Phi}\left(k_{1},k_{2}\right)\right|^{2}=1$, it can be shown that
\begin{equation}\label{eq:Q_0_second}
Q_0=\sqrt{\frac{2}{\pi\sigma_{+}\sigma_{-}}}.
\end{equation}
In Fig.~\ref{fig:biphoton}, we plot three sample biphoton wave functions of the form given in Eq.~\eqref{biphoton_form} for different $\sigma_+$ and $\sigma_-$ for $k_0=\pi/2D$.\par
\begin{figure}[ht]
\begin{center}
\includegraphics[scale=0.53]{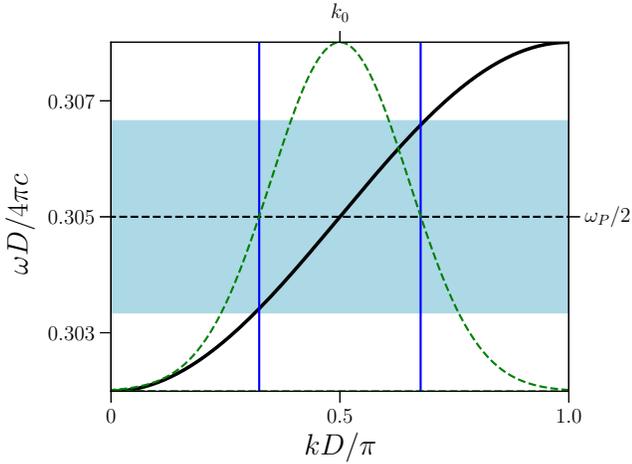}
\caption{The CROW dispersion relation and biphoton wave function. The solid black line shows the frequency as a function of the Bloch vector for the CROW structure. The dashed horizontal line gives the pump frequency divided by two. The dashed green line gives the function $\Phi(k,-k)$ for $k_0=\pi/(2D)$. The two vertical solid blue lines indicate the FWHM in $k$, while the shaded blue region indicates the FWHM in frequency, both of which can be found from Eq.~\eqref{biphoton_02} for $\sigma_+D=\sigma_-D=0.47$.} \label{fig:dispersion}
\end{center}
\end{figure}

 
To graphically illustrate the validity of the assumptions made to obtain Eq.~(\ref{biphoton_form}), in Fig.~\ref{fig:dispersion} we plot the dispersion of our CROW. The physical parameters of the CROW are from Ref.~\cite{doi:10.1063/1.2737430}. It consists of a dielectric slab of refractive index $n=3.4$ having a square array of cylindrical air voids of radius $a=0.4d$, height $h=0.8d$, and lattice vectors $\textbf{a}_1=d\hat{\textbf{x}}$ and $\textbf{a}_2=d\hat{\textbf{y}}$, where $d$ is the period. The cavities are point defects formed by periodically removing air voids in a line with $D=2d$ (see Fig.~\ref{fig:schematic}). The complex frequency, $\tilde{\omega}_F$, and the complex coupling parameter, $\tilde{\beta}_1$, of the structure are $(0.305-i7.71\times10^{-6})4\pi c/D$, and $9.87\times10^{-3}-i1.97\times10^{-5}$, respectively. To visualize the biphoton wave function superimposed on the CROW dispersion, we plot ${\Phi}(k,-k)$ for $W_S=3D$ and $\sigma_+=\sigma_-$ in Fig.~\ref{fig:dispersion} as well. As can be seen, the first order expansion of $\cos(k_1D)$ and $\cos(k_2D)$ about $k_0$ and $-k_0$ is accurate as the dispersion within this range is very close to linear. In order for our Schmidt decomposition to be valid for this structure, $\sigma_+D$ and $\sigma_-D$ cannot be increased significantly beyond the chosen value of $0.47$ otherwise the biphoton wave function will not be confined to the quadrant where $k_1>0$ and $k_2<0$. Note that increasing the pump width to higher values, $W_S>3D$, increases the accuracy of our approximation, as is evident from Figs.~\ref{fig:biphoton}(b) and \ref{fig:biphoton}(c) where $W_S$ is $5.05D$ and $10.10D$, respectively. \par  
Before employing a Schmidt decomposition, we present the relation between $\sigma_\pm$ and the temporal and spatial full width at half maximum (FWHM) of the pump pulse. Using Eqs.~(\ref{eq:f_yz_text}) and (\ref{eq:sigma_minus}), the \textit{temporal} FWHM of the pump can be written as
\begin{equation}\label{eq:delta_t}
\Delta t_{FWHM}=\frac{2\sqrt{\ln2}\tau}{\sigma_{-}D},
\end{equation}   
where $\tau\equiv1/Re(\tilde{\omega}_F\tilde{\beta}_1)$ is the time for a pulse with Bloch vector $k=k_0=\pi/(2D)$ to travel one period. Similarly, using Eqs.~(\ref{PHI_qx_text}) and (\ref{eq:sigma_plus}), the \textit{spatial} FWHM of the pump is found to be
\begin{equation}\label{eq:deltar}
\Delta r_{FWHM}=\frac{2\sqrt{\ln2}}{\sigma_{+}}.
\end{equation}
Using these two equations, one can obtain a clear understanding of the necessary pumping conditions. For instance, the quantities considered in Fig.~\ref{fig:biphoton}(a) correspond to a pump with $3.54\tau$ and $3.54D$ as the temporal and spatial FWHM, respectively. As we show later, for our CROW, the propagation loss in the system is very small while the squeezed light is being generated, which validates the neglect of loss during the generation process.   \par
The special form of the biphoton wave function, Eq.~(\ref{biphoton_form}), allows us to perform a Schmidt decomposition analytically~\cite{U'Ren:2003:PEQ:2011564.2011567,doi:10.1080/09500340600777805,PhysRevA.96.053842} for $\sigma_{-}\geq\sigma_{+}$ as
\begin{equation}
\label{Schmidt}
\begin{multlined}
\sqrt{\frac{2}{\pi\sigma_{+}\sigma_{-}}}\exp\left(-\frac{(\delta_{1}+\delta_{2})^{2}}{2\sigma_{+}^{2}}\right)\exp\left(-\frac{\left(\delta_{1}-\delta_{2}\right)^{2}}{2\sigma_{-}^{2}}\right)\\
=\sum_{\lambda}\sqrt{p_{\lambda}}\psi_{\lambda}\left(\delta_{1}\right)\psi_{\lambda}\left(\delta_{2}\right),
\end{multlined}
\end{equation}
where
\begin{equation}
\label{p_lambda}
p_{\lambda}=4\sigma_{+}\sigma_{-}\frac{\left(\sigma_{+}-\sigma_{-}\right)^{2\lambda}}{\left(\sigma_{+}+\sigma_{-}\right)^{2\left(\lambda+1\right)}},
\end{equation}
\begin{equation}\label{eq:psi_lambda}
\begin{multlined}
\psi_{\lambda}\left(\delta\right)=\left(-i\right)^{\lambda}\sqrt{\frac{\sqrt{2}}{2^{\lambda}\lambda!\sqrt{\pi\sigma_{+}\sigma_{-}}}}\\
\times\exp\left(-\frac{\delta^{2}}{\sigma_{+}\sigma_{-}}\right)H_{\lambda}\left(\frac{\sqrt{2}\delta}{\sqrt{\sigma_{+}\sigma_{-}}}\right),
\end{multlined}
\end{equation}
and the $H_{\lambda}\left(x\right)$ are Hermite polynomials of order $\lambda$. Note that the Schmidt number is given by~\cite{Zhukovsky:12}
\begin{equation}
K=\frac{1}{\sum_{\lambda}p_{\lambda}^{2}}=\frac{\sigma_{+}^{2}+\sigma_{-}^{2}}{2\sigma_{+}\sigma_{-}}.
\end{equation} \par
\begin{figure}[!htb]
\includegraphics[scale=0.54]{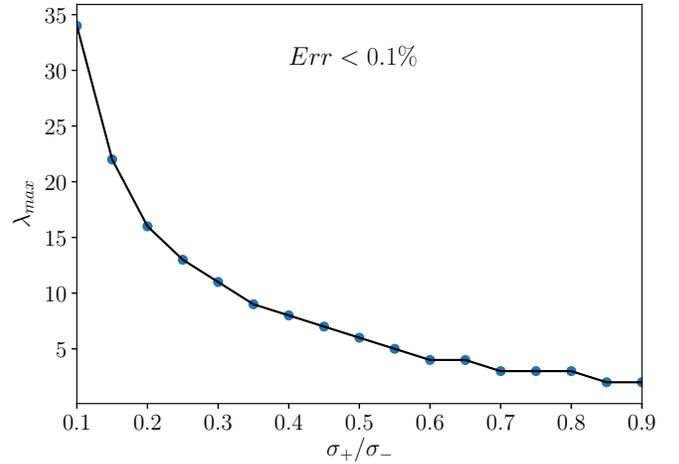}
\caption{Maximum number of Schmidt modes required to be considered in $\Phi$ as a function of $\sigma_{+}/\sigma_{-}$ to ensure $Err < 0.1\%$.  }\label{fig:max_lambda}
\end{figure}

Using Eq.~(\ref{p_lambda}) for the special case where $ \sigma_-=\sigma_+ =\sigma$, it can be shown that the only nonzero term in Eq.~(\ref{ph_number_01}) is for $ \lambda=0 $. However, in general, one needs to include several of the Schmidt modes (up to $\lambda=\lambda_{max}$) to accurately represent the biphoton wave function. To quantify the accuracy of the Schmidt decomposition used in our calculations, we define an error function as
\begin{equation}
Err=\sqrt{\frac{\int\text{d}k_{1}\,\text{d}k_{2}\,|{\Phi}\left(k_1,k_2\right)-{\Phi_{App}}\left(k_1,k_2\right)|^2}{\int\text{d}k_{1}\,\text{d}k_{2}\,|{\Phi}(k_1,k_2)|^2}},
\end{equation}  
where {$\Phi$ and $\Phi_{App}$} are the exact and approximate expressions, respectively, given by Eq.~(\ref{biphoton_form}) and 
\begin{equation}
 {\Phi_{App}(k_0+\delta_1,-k_0+\delta_2)}=\sum_{\lambda=0}^{\lambda_{max}}\sqrt{p_{\lambda}}\psi_{\lambda}\left(\delta_{1}\right)\psi_{\lambda}\left(\delta_{2}\right).
\end{equation}
In Fig.~\ref{fig:max_lambda} we plot the index of the maximum Schmidt mode needed to be included in ${\Phi_{App}}$, in order to ensure $Err < 0.1\%$. For example, as can be seen, when $\sigma_{-}=2\sigma_{+}$, one needs to include $6$ terms~($\lambda_{max}=6$) to achieve the desired accuracy.\par
Using these results in Eqs.~(\ref{eq:apdap}) and (\ref{eq:bdkbk}), the time-dependent average photon number in the $ p^{th} $ cavity is found to be
\begin{widetext}
\begin{equation}
\begin{multlined}
\label{ph_number_01}
\left\langle \hat{a}_{p}^{\dagger}\left(t\right)\hat{a}_{p}\left(t\right)\right\rangle         =       \frac{D}{2\pi}e^{-2Re\left(i\tilde{\omega}_{F}\left(1-\tilde{\beta}_{1}\cos(|k_{0}|D)\right)t\right)}\\\times\sum_{\lambda}\sinh^{2}\left(r_\lambda\right)\frac{\sqrt{2\sigma_{+}\sigma_{-}\pi}}{2^{\lambda}\lambda!}\times    \left(\left|H_{\lambda}\left(\tilde{S}_{p-}\right)\right|^{2}e^{-\left(\frac{\left(\tilde{S}_{p-}^{*2}+\tilde{S}_{p-}^{2}\right)}{2}\right)}
        +\left|H_{\lambda}\left(\tilde{S}_{p+}\right)\right|^{2}e^{-\left(\frac{\left(\tilde{S}_{p+}^{*2}+\tilde{S}_{p+}^{2}\right)}{2}\right)}\right),
\end{multlined}
\end{equation}
\end{widetext}
where 
\begin{align}
\tilde{S}_{p\pm}=\left(\tilde{\omega}_{F}\tilde{\beta}_{1}D\sin\left(|k_{0}|D\right)t\pm pD\right)\sqrt{\frac{\sigma_{+}\sigma_{-}}{2}}. 
\end{align}
The time-dependent average photon number in the $ p^{th} $ cavity when $\sigma_{+}=\sigma_{-}$ can be simplified to 
\begin{equation}
\begin{split}
\label{ph_number_02}
\left\langle \hat{a}_{p}^{\dagger}\left(t\right)\hat{a}_{p}\left(t\right)\right\rangle         &=       \frac{\sigma D}{\sqrt{2\pi}}e^{-2Re\left(i\tilde{\omega}_{F}\left(1-\tilde{\beta}_{1}\cos(|k_{0}|D)\right)t\right)}\sinh^{2}\left(r_{0}\right)\\
&\times\left(e^{-\left(\frac{\left(\tilde{S}_{p-}^{*2}+\tilde{S}_{p-}^{2}\right)}{2}\right)}
        +e^{-\left(\frac{\left(\tilde{S}_{p+}^{*2}+\tilde{S}_{p+}^{2}\right)}{2}\right)}\right).
\end{split}
\end{equation}
We note that using Eq.~(\ref{ph_number_02}) for a lossless system, one finds that the total number of photons in the CROW is $2\sinh^2\left(r_{0}\right)$, independent of $\sigma$, which agrees with the total number of generated photons in any two-mode squeezed state.  \par
Following a procedure similar to that used to arrive at Eq.~(\ref{ph_number_01}), one can derive the following expectation value for $\left\langle \hat{a}_{p}(t)\hat{a}_{p^{\prime}}(t)\right\rangle$, which is needed in Eq.~(\ref{eq:entanglement_01}) to evaluate the variances of the quadrature operators and the correlation variance in the CROW structure: 
\begin{widetext}
\begin{equation}
\begin{multlined}
\left\langle \hat{a}_{p}(t)\hat{a}_{p^{\prime}}(t)\right\rangle =\frac{D}{2\pi}\sum_{\lambda}(-1)^{\lambda}\left(\cosh(r_\lambda)\sinh\left(r_\lambda\right)\right)e^{-i2\tilde{\omega}_{F}\left(1-\tilde{\beta}_{1}\cos(|k_{0}|D)\right)t}\frac{\sqrt{2\sigma_{+}\sigma_{-}\pi}}{2^{\lambda}\lambda!}\\
\times\left[e^{-i|k_{0}|\left(p-p^{\prime}\right)D}H_{\lambda}\left(\tilde{S}_{p+}\right)H_{\lambda}\left(\tilde{S}_{p^{\prime}-}\right)e^{-\left(\frac{\left(\tilde{S}_{p+}^{2}+\tilde{S}_{p^{\prime}-}^{2}\right)}{2}\right)}\right.+\left.e^{i|k_{0}|\left(p-p^{\prime}\right)D}H_{\lambda}\left(\tilde{S}_{p-}\right)H_{\lambda}\left(\tilde{S}_{p^{\prime}+}\right)e^{-\left(\frac{\left(\tilde{S}_{p-}^{2}+\tilde{S}_{p^{\prime}+}^{2}\right)}{2}\right)}\right].
\end{multlined}\label{Eq:apapp}
\end{equation}
\end{widetext}
Note that $\left\langle \hat{a}_{p}^{\dagger}(t)\hat{a}_{p^{\prime}}^{\dagger}(t)\right\rangle$ is simply the complex conjugate of Eq.~(\ref{Eq:apapp}).\par
In the next section, we will use these equations to determine the photon number and correlation variance under a variety of different pump conditions.  
\section{\label{sec:level4}Results}
\begin{figure}
\begin{center}
\includegraphics[scale=0.42]{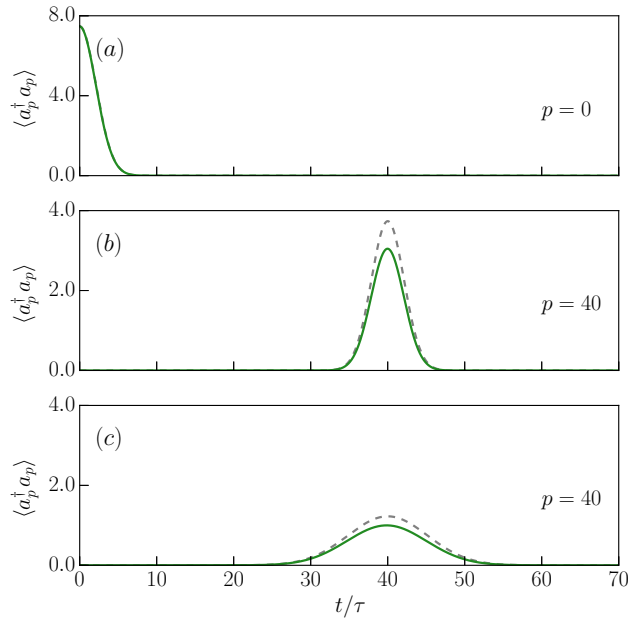}
\caption{Average photon number in the (a) central and (b) fortieth cavities of the CROW as a function of time for $\sigma_+\,D=\sigma_-\,D=0.47$ and $\beta=2.2$. (c) The time dependent average photon number for $\sigma_+\,D=0.14$, $\sigma_-\,D=0.28$, and $\beta=2.2$. The dashed grey lines show the case in which the effect of loss is ignored. }\label{fig:ph_number}
\end{center}
\end{figure}

Using the expectation values derived in Sec.~\ref{sec:result_for_gaussian}, we can now study the photon evolution and inseparability criteria for the generalized two-mode squeezed light inside the CROW structure. In Fig.~\ref{fig:ph_number}, we plot the average number of photons in the $p$th cavity for $p=0,40$ as a function of time for both a lossy~(solid green line) and lossless~(dashed grey line) system. The propagation of light between the coupled cavities and effect of loss on the number of photons in each cavity is evident in Figs.~\ref{fig:ph_number}(a) and \ref{fig:ph_number}(b), where $\sigma_+\,D=\sigma_-\,D=0.47$. Because the system and pump are spatially symmetric, the results are identical for $p\rightarrow-p$. In Fig.~\ref{fig:ph_number}(c), we plot the time-dependent average photon number in the $40$th cavity with $\beta=2.2$ still, but with the $\sigma_+\,D=0.14$ and $\sigma_-\,D=0.28$. As can be seen, the pulse width is wider for this smaller $\sigma_+$, as expected. \par
Since our input pump state is a coherent state, the number of pump photons is $N_P=|\alpha|^2$. We now examine what the pump parameters will be for a specific case of interest. We again consider the case where $\sigma_{+}D=\sigma_{-}D=0.47$; using Eqs.~(\ref{eq:delta_t}) and (\ref{eq:deltar}) this gives temporal and spatial FWHM for the pump of $295\,\text{fs}$ and $3.3\,\mu \text{m}$, respectively. We choose the CROW material to be $\text{Al}_{0.35}\text{Ga}_{0.65}\text{As}$ due to its high nonlinearity and relatively large bandgap. In addition, we choose the pump wavelength to be $\lambda_S=775\,\text{nm}$, which not only results in generating counterpropagating signal and idler photons at the telecommunication wavelength, $\lambda_F=1550\,\text{nm}$, but also ensures operation below the band gap of $\text{Al}_{0.35}\text{Ga}_{0.65}\text{As}$. Choosing the periodicity of the CROW structure to yield a signal central wavelength of $1550\,\text{nm}$, gives $D\approx0.9\,\mu \text{m}$. Using Eq.~(\ref{eq:Xi2bar}) and the normalization condition given in Appendix \ref{appendix_normalization},  $\bar{\chi}_2$ for our structure is approximately given by $\bar{\chi}_2\approx\chi_2/n^2(\omega_F)$, where $\chi_2\approx 100~\text{pm/V}$, appropriate for AlGaAs alloys~\cite{Yang:07,Gili:16,Carletti:15}, and $n\approx3.4$ at $\omega_{F}$. We now seek to determine the approximate number of pump photons under the above conditions that will give a squeezing parameter of $2.2$. Employing Eqs.~(\ref{eq:Q_0}) and (\ref{eq:Q_0_second}), the average number of photons in the pump is found to be $7.4\times10^{10}$, which gives a total pump pulse energy of approximately $19~\text{nJ}$.  We note that all of the above pump characteristics are easily achievable from a Ti:Sapphire laser.    \par 
\begin{figure}
\begin{center}
\includegraphics[scale=0.42]{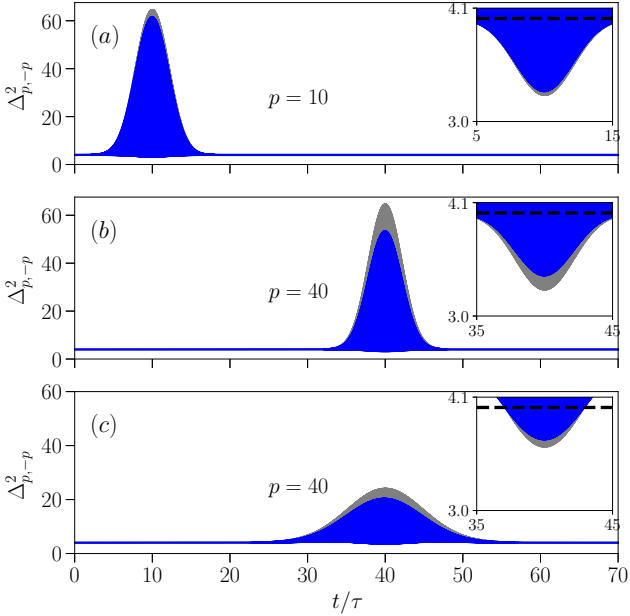}
\caption{(a-b) Correlation variance between different pairs of cavities in the CROW as a function of time for $\sigma_+\,D=\sigma_-\,D=0.47$ and $\beta=2.2$. (c) The time dependent correlation variance for $\sigma_+\,D=0.14$, $\sigma_-\,D=0.28$, and $\beta=2.2$. The results for a lossless system are shown in gray. }\label{fig:Entanglement}
\end{center}
\end{figure}

In Fig.~\ref{fig:Entanglement} we plot the time-dependent correlation variances for different sets of lossy and lossless cavities in blue and grey, respectively. Note that there are fast oscillations that are not observable on this time scale. The dashed lines in the insets show the inseparability criteria below which the light is considered to be entangled. Here we only focus on cases where the two cavities considered are located the same distance from the central cavity, as this will yield the maximum entanglement; however, using Eq.~(\ref{eq:entanglement_01}) one can explore the entanglement between any two cavities of the CROW. As can be seen in Fig.~\ref{fig:Entanglement}(a) and (b), due to the loss in the system, the degree of entanglement decreases as the system evolves in time, whereas for a lossless system (the grey color) the degree of entanglement does not change as the light propagates.
\par
\begin{figure}[!hbt]
\begin{center}
\includegraphics[scale=0.52]{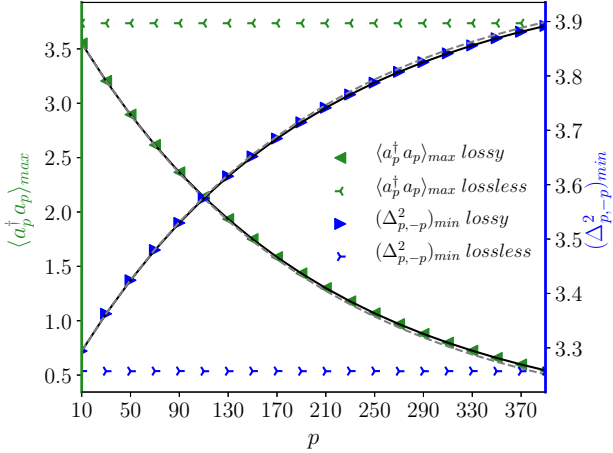}\
\caption{Maximum number of photons (left axis) as a function of the cavity index and minimum correlation variances (right axis) between different symmetrically displaced pairs of cavities for $\sigma_+\,D=\sigma_-\,D=0.47$ and $\beta=2.2$. The solid black and dashed grey lines represent the results from Eqs.~(\ref{eq:ph_num_max_p}) and (\ref{eq:ent_max_p}) with and without including $\exp[{\left(V_{I}t_{max}\right)^{2}\frac{\sigma^{2}}{2}}]$, respectively.   }\label{fig:maximums}
\end{center}
\end{figure}  
In Fig.~\ref{fig:maximums} we plot the maximum number of photons for a lossless and lossy CROW as a function of the cavity index, $p$. As expected, when loss is included, the number of photons decreases as we move away from the central cavity. In Fig.~\ref{fig:maximums} we also plot the minimum correlation variance for the lossless and lossy CROW as a function of cavity index, $p$. As can be seen, due to the reduction in the number of photons in the lossy case, there is a decrease in the degree of entanglement as a function of $p$. For instance, the minimum correlation variance at the tenth cavity is $0.9$ times the corresponding value at the one hundredth cavity. \par 

For a general pump, the evolution equations for the photon number and correlation variance are quite complicated and it is difficult to discern the general behaviour or the effects of loss from the full equations. However, for the special case where $\sigma_+=\sigma_{-}=\sigma$ and $k_0=\pi/2D$, approximate analytic expressions can be obtained. We begin by defining the complex quantity, $\tilde{V}=V_{R}+iV_{I}=\tilde{\omega}_{F}\tilde{\beta}_{1}D$, which enables us to rewrite $\tilde{S}_{p\pm}^{2}$ as 
\begin{equation}
\begin{multlined}
\tilde{S}^2_{p\pm}=\left(\left(V_{R}+iV_{I}\right)t\pm pD\right)^{2}\frac{\sigma^{2}}{2}\\=\left(\left(V_{R}t\pm pD\right)^{2}+2i\left(V_{R}t\pm pD\right)V_{I}t-\left(V_{I}t\right)^{2}\right)\frac{\sigma^{2}}{2}.
\end{multlined}
\end{equation} 
Considering only the dominant terms in Eq.~(\ref{ph_number_02}), to a very good approximation one can show that the time at which the photon number in the $p^{th}$ cavity peaks in a lossless system is $t_{max}=p\tau\approx p/\omega_{F}\beta_{1}$. As can be seen in Fig.~\ref{fig:ph_number}, the photon number peaks at essentially the same time in both lossy and lossless system. Using $t_{max}$, we are able to derive the following approximate expression for the maximum photon number in the $p^{th}$ cavity (for $p>0$) in a lossy system:
\begin{align}
\left\langle a_{p}^{\dagger}a_{p}\right\rangle_{max} \approx \frac{\sigma D}{\sqrt{2\pi}}\sinh^{2}\left(r_{0}\right)e^{-2\gamma_{F}p\tau}e^{\left(V_{I}p\tau\right)^{2}\frac{\sigma^{2}}{2}}.
\label{eq:ph_num_max_p}
\end{align}
Following the same procedure and using Eqs.~(\ref{eq:entanglement_01}) and (\ref{Eq:apapp}), one obtains
\begin{align}
\left(\Delta_{p,-p}^{2}\right)_{min}\approx & 4+4\frac{\sigma D}{\sqrt{2\pi}}\left(2\sinh^{2}\left(r_{0}\right)-\sinh\left(2r_{0}\right)\right)\nonumber\\
&\quad\times e^{-2\gamma_{F}p\tau}e^{\left(V_{I}p\tau\right)^{2}\frac{\sigma^{2}}{2}}.
\label{eq:ent_max_p}
\end{align}

We note first that both the photon number and the deviation of the correlation variance from $4$ depend linearly on $\sigma$.  Thus, as expected, the separability is largest when the pump is short in time and narrow in space (for a fixed squeezing parameter, $r_{0}$). Now, according to Eqs.~(\ref{eq:ph_num_max_p}) and (\ref{eq:ent_max_p}), under the above-mentioned pumping conditions the effect of loss on the maximum number of photons and on the entanglement as a function of $p$ is given by two exponential factors. The first factor accounts for the intrinsic loss in an individual cavity (which is also the intrinsic loss of the Bloch mode with $k=\pi / 2D$).  The second factor accounts for the loss dispersion in the CROW, and results in a reduction in the loss. Of course, these analytic results are only valid when the effect of the dispersion of the loss is small.

We now consider how well these approximate expressions reproduce the exact results.  In Fig.~\ref{fig:maximums}, we plot the results of Eqs.~(\ref{eq:ph_num_max_p}) and (\ref{eq:ent_max_p}) with (black solid line) and without (red solid line) including the factor, $\exp[{\left(V_{I}p\tau\right)^{2}\frac{\sigma^{2}}{2}}]$. As can be seen, for this CROW the full approximate analytic expressions very accurately reproduces the exact results.  Moreover, to a very good approximation, one can evaluate Eqs.~(\ref{eq:ph_num_max_p}) and (\ref{eq:ent_max_p}) neglecting the loss-dispersion factor, $\exp[{\left(V_{I}p\tau\right)^{2}\frac{\sigma^{2}}{2}}]$. For example, for $\sigma_-D$=$\sigma_+D=0.47$ and $p=350$, the first and the second exponential factors in Eqs.~(\ref{eq:ph_num_max_p}) and (\ref{eq:ent_max_p}) are $0.18$ and $1.10$, respectively, showing that loss dispersion only changes the results by $10\%$. In general, it can be shown that in order to have less than $10\%$ error in evaluating the photon number and the difference of the correlation variance from $4$, the range of $p$ must be limited to $p\le\sqrt{2}/(10V_I\tau\sigma)$.  \par     
Finally, we now consider the more general cases in which $\sigma_{-}$ is not necessarily equal to $\sigma_{+}$. Under these conditions, we cannot derive simple expressions for the maximum photon number and entanglement as a function of $p$. We present the results of the full calculations for a \textit{lossless} system in Table~\ref{table1} for a number of different pump durations and spatial widths; all other parameters are the same as in the previous plots. As can be seen, the maximum entanglement is obtained when the pump duration is as short as possible and the pump width is as narrow as possible (i.e. $\sigma_+D=\sigma_-D=0.47$ for our system). \par
\begin{figure}[!h]
\begin{center}
\includegraphics[scale=0.52]{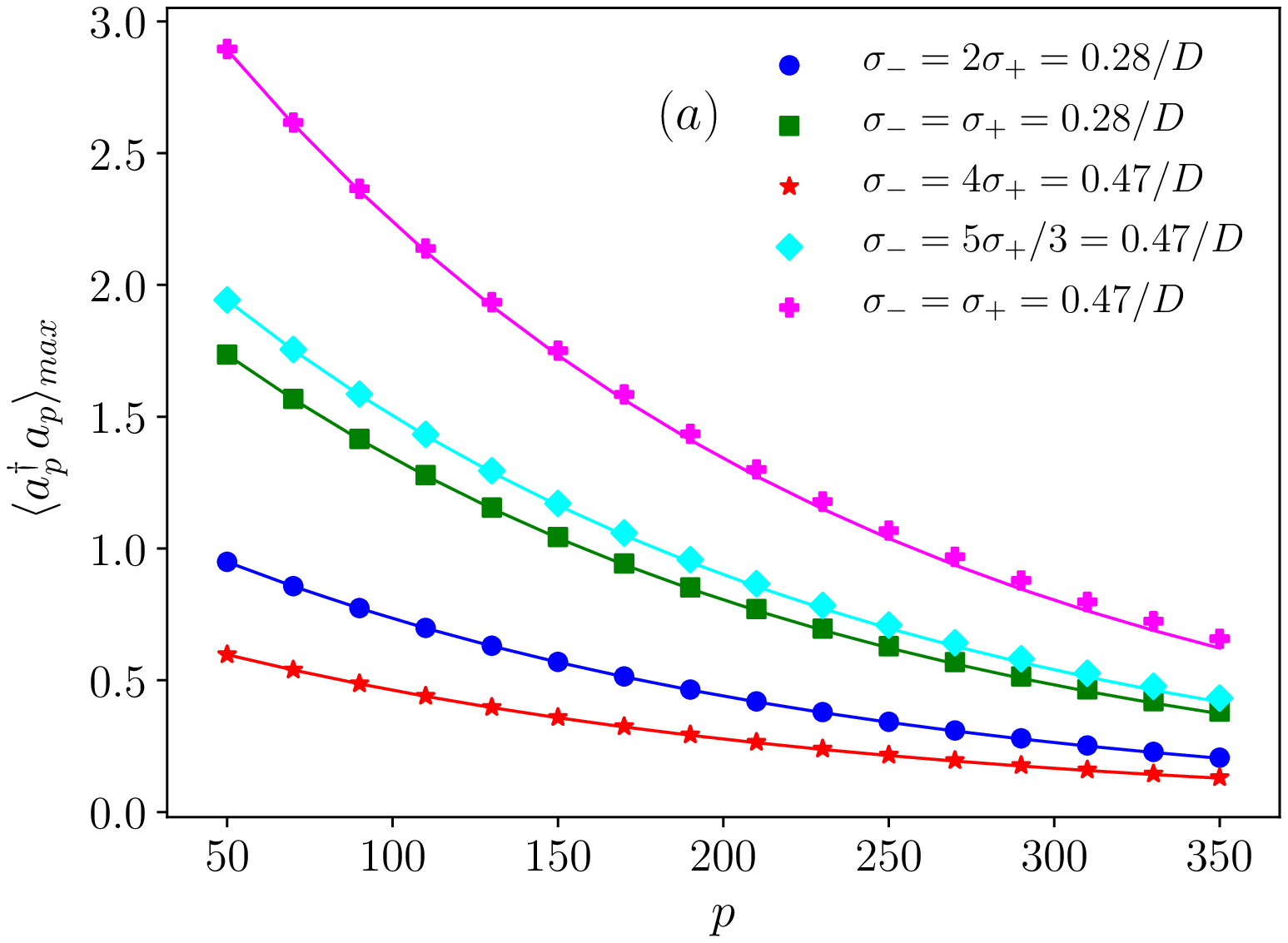}\
\includegraphics[scale=0.52]{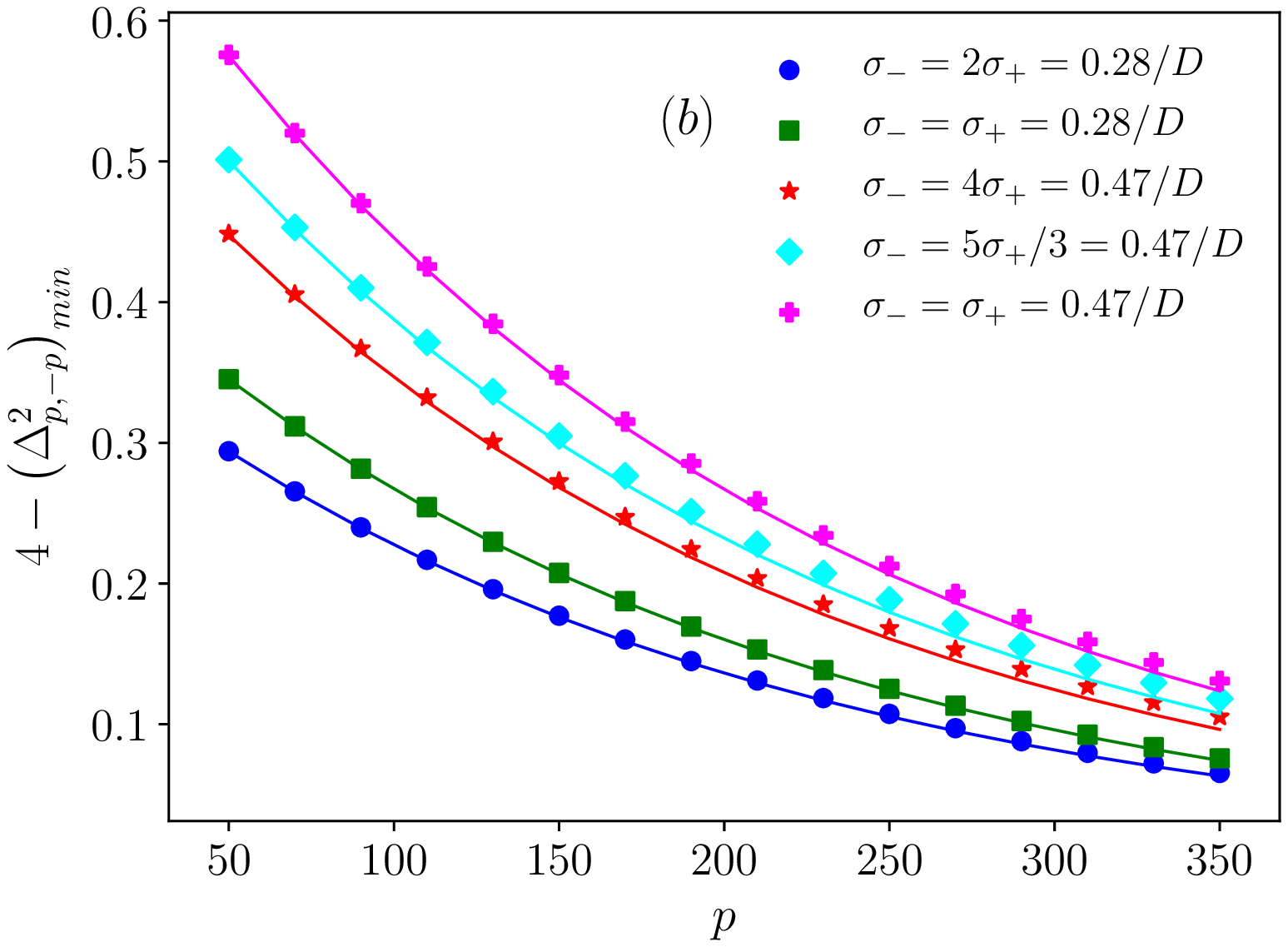}\
\caption{(a) Maximum number of photons and (b) the deviation of the minimum of the correlation variance from the inseparability threshold of $4$ for a lossy system and selected range of cavities for $\beta=2.2$ and different pumping configurations. The solid lines represent the results from data given in Table.~\ref{table1} and the exponential factor $\exp{(-2\gamma_{F}p\tau)}$, and the markers show the results from the full calculations.   }\label{fig:maximums_not_equal_sigma_appendix}
\end{center}
\end{figure}
\begin{figure}[!htb]
\begin{center}
\includegraphics[scale=0.52]{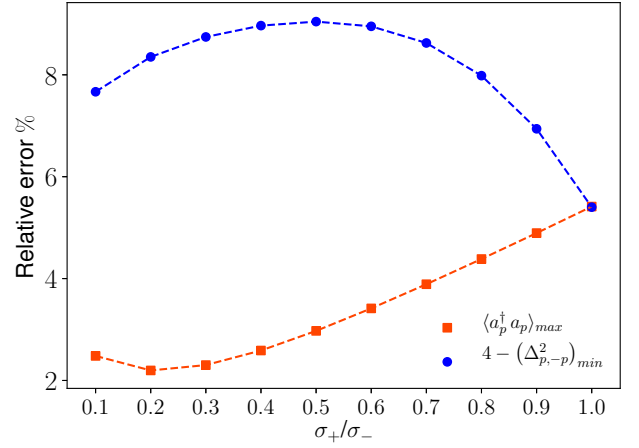}
\caption{Relative deviation of the results by the exponential factor $\exp{(-2\gamma_{F}p\tau)}$ from the results of the full calculation as a function of $\sigma_+/\sigma_-$ for $p=350$ and $\sigma_-D=0.47$.}
\label{fig:Frac_diff_350}
\end{center}
\end{figure}

We have also performed full calculations for the evolution in the presence of loss for different pump configurations. In Fig.~\ref{fig:maximums_not_equal_sigma_appendix}, we compare the results of the full calculations with the results when the loss is incorporated approximately using only the exponential factor $\exp{(-2\gamma_{F}p\tau)}$. We plot the maximum number of photons and the deviation of the minimum of the correlation variance from the inseparability threshold of 4 for a lossy system as a function of $p$ for different pumping configurations. As can be seen, for short distances from the central cavity~(small $p$), the effect of loss on the result can, to a very good approximation, still be explained by only including the exponential factor $\exp{(-2\gamma_{F}p\tau)}$. However, for the cavities far from the central cavity (large $p$), although the general trends can still be predicted by such an approximation, the difference between the exact and approximation results becomes pronounced and the validity of this approximation becomes questionable. In order to show the difference between exact and approximate results for large $p$, in Fig.~\ref{fig:Frac_diff_350} we plot the relative difference between the results from the full calculations and those from approximating the loss in the system by only considering the exponential factor $\exp{(-2\gamma_{F}p\tau)}$ as a function of $\sigma_+/\sigma_-$, for $p=350$ and $\sigma_-D=0.47$. 
As expected, when $\sigma_+=\sigma_-$, we obtain a relative error of approximately $5.4\%$, which is simply due to the dispersion factor, $\exp[{\left(V_{I}p\tau\right)^{2}\frac{\sigma^{2}}{2}}]$.  However, in general, the error depends on $\sigma_+/\sigma_-$, and is different for the photon number and correlation variance, due to the different way in which loss dispersion affects these two quantities.
As can be seen in this specific example, evaluating $4-\left(\Delta_{p,-p}^{2}\right)_{min}$ using the approximation method results in a $9 \%$ deviation from the results of the full calculations when $\sigma_{+}=0.5\sigma_{-}$, while, the relative difference for the maximum number of photons is always less than $5.4\%$. It is thus evident that a simple exponential factor will capture the general effect of loss on the correlation, but it may not be very accurate depending on the structure and the pump conditions.

\setlength{\tabcolsep}{0.9em}
\begin{table}
\centering
\caption{The maximum number of photons and the deviation of the minimum of the correlation variance from the inseparability threshold of $4$ in a lossless system for $\beta=2.2$ and different pumping configurations}
\label{table1}
 \begin{tabular}{c  c  c  c  c  c} 
 \hline\
   \makecell{$\sigma_-D$\\$\sigma_+D$}& \makecell{$0.28$\\$0.14$} & \makecell{$0.28$\\$0.28$} & \makecell{$0.47$\\$0.12$} &\makecell{$0.47$\\$0.28$} & \makecell{$0.47$\\$0.47$} \\ [1ex] 
 \hline\\
 $\left\langle a_{p}^{\dagger}a_{p}\right\rangle_{max}$& 1.23 & 2.24 & 0.77 & 2.51 & 3.74  \\ [1ex]
 $4-\left(\Delta_{p,-p}^{2}\right)_{min}$  & 0.38 & 0.44 & 0.58 & 0.65 & 0.74  \\
 [1ex]
 \hline
 \end{tabular}
\end{table}
\section{Conclusion}\label{sec:conclusion}
In this work we studied the generation and propagation of entangled states in lossy coupled-cavity systems. We applied the tight-binding method to evaluate the fields and complex frequencies for the leaky modes of lossy coupled-cavity system, and presented analytic time-dependent expressions for the photon number and correlation variance in a lossy CROW structure. We showed how properties such as the average number of photons in each cavity and the correlations between cavities are affected by loss. For the CROW structure considered in this work, we found that as the light gets far from the central cavity, the effects of loss become more significant and cannot be ignored. Moreover, we obtained  simple, approximate analytic expressions for the effects of loss on the propagation of the generated light in the CROW, and have shown that they can be used to predict general trends. However, depending on the details of the pumping conditions and the CROW structure itself, the accuracy of this approximation varies, and to get an accurate result, specifically for cavities far from the central cavity, the effect of loss cannot be well-described using a simple exponential factor that is given by the loss in an individual cavity. Using full numerical results is suggested for optimization. Yet these analytic results allow researchers to easily explore the spectral and spatial pumping configurations needed to generate the counterpropagating entangled states in a CROW.   
\appendix

\section{}
\label{appendix_normalization}
In this appendix we present the normalization condition for the modes. Considering the refractive index of the material to be nondispersive within the range of frequency considered here, following Yang \textit{et al}.~\cite{PhysRevA.77.033808}, the normalization conditions can be written as
\begin{equation}
\int\text{d}\mathbf{r}\frac{\mathbf{D}^*_{Fk}(\mathbf{r})\cdot\mathbf{D}_{Fk^\prime}(\mathbf{r})}{\varepsilon_0 n^2(\mathbf{r};\omega_{F})}=\delta(k-k^\prime)
\label{eq:D_norm}
\end{equation}
and
\begin{equation}
\int\text{d}\mathbf{r}\frac{\mathbf{M}^*_{Sm\mathbf{q}}(\mathbf{r})\cdot\mathbf{M}_{Sm^\prime\mathbf{q^\prime}}(\mathbf{r})}{\varepsilon_0n^{2}\left(\mathbf{r};\omega_{Sm}\right)}=\delta_{mm^\prime}\delta(\mathbf{q}-\mathbf{q^\prime}).
\end{equation}   
Using Eq.~(\ref{eq:D_in terms of N}) in Eq.~(\ref{eq:D_norm}) we obtain the normalization condition for the single-cavity modes as
\begin{equation}
\int\text{d}\mathbf{r}\frac{N^*_{Fp}(\mathbf{r})\cdot N_{Fp^\prime}(\mathbf{r})}{\varepsilon_0 n^2(\mathbf{r};\omega_F)}=\delta_{pp^\prime}.
\end{equation}

\section{}
\label{appendix_B}
In this appendix we present an accurate approximate analytic result for the summation in Eq.~(\ref{eq:biphoton_sigma_and_integral}).
Starting from Eq.~(\ref{eq:biphoton_sigma_and_integral}) we have
\begin{equation}
S=\sum_{p}e^{-\frac{p^{2}D^{2}}{W_S^{2}}}e^{-ip(k_{1}+k_{2})D}.		
\end{equation}
Approximating this sum as an integral, we obtain
\begin{equation}
S\approx\int dp\,e^{-\frac{p^{2}D^{2}}{W_S^{2}}}e^{-ip(k_{1}+k_{2})D}
	=	\frac{1}{D}\int dp^{\prime}\,e^{-\frac{p^{\prime2}}{W_S^{2}}}e^{-i2\pi p^{\prime}k^{\prime}},
\end{equation}
where $p^{\prime}=pD$ and $k^{\prime}=\frac{k_{1}+k_{2}}{2\pi}$. Now using 
\begin{equation}
\int_{-\infty}^{\infty}e^{-ax^{2}}e^{-2\pi ikx}dx=\sqrt{\frac{\pi}{a}}e^{-\frac{\pi^{2}k^{2}}{a}},
\end{equation}
we can write
\begin{equation}
S=\frac{W_S\sqrt{\pi}}{D}\,\exp\left(-\left(\frac{\left(k_{1}+k_2\right)W_S}{2}\right)^{2}\right).
\end{equation}
One can show numerically that for $W_S\geq2D$, the approximation made here is accurate to within $0.01 \%.$ 

\clearpage
\bibliography{mybib}

\begin{thebibliography}{48}%
\makeatletter
\providecommand \@ifxundefined [1]{%
 \@ifx{#1\undefined}
}%
\providecommand \@ifnum [1]{%
 \ifnum #1\expandafter \@firstoftwo
 \else \expandafter \@secondoftwo
 \fi
}%
\providecommand \@ifx [1]{%
 \ifx #1\expandafter \@firstoftwo
 \else \expandafter \@secondoftwo
 \fi
}%
\providecommand \natexlab [1]{#1}%
\providecommand \enquote  [1]{``#1''}%
\providecommand \bibnamefont  [1]{#1}%
\providecommand \bibfnamefont [1]{#1}%
\providecommand \citenamefont [1]{#1}%
\providecommand \href@noop [0]{\@secondoftwo}%
\providecommand \href [0]{\begingroup \@sanitize@url \@href}%
\providecommand \@href[1]{\@@startlink{#1}\@@href}%
\providecommand \@@href[1]{\endgroup#1\@@endlink}%
\providecommand \@sanitize@url [0]{\catcode `\\12\catcode `\$12\catcode
  `\&12\catcode `\#12\catcode `\^12\catcode `\_12\catcode `\%12\relax}%
\providecommand \@@startlink[1]{}%
\providecommand \@@endlink[0]{}%
\providecommand \url  [0]{\begingroup\@sanitize@url \@url }%
\providecommand \@url [1]{\endgroup\@href {#1}{\urlprefix }}%
\providecommand \urlprefix  [0]{URL }%
\providecommand \Eprint [0]{\href }%
\providecommand \doibase [0]{http://dx.doi.org/}%
\providecommand \selectlanguage [0]{\@gobble}%
\providecommand \bibinfo  [0]{\@secondoftwo}%
\providecommand \bibfield  [0]{\@secondoftwo}%
\providecommand \translation [1]{[#1]}%
\providecommand \BibitemOpen [0]{}%
\providecommand \bibitemStop [0]{}%
\providecommand \bibitemNoStop [0]{.\EOS\space}%
\providecommand \EOS [0]{\spacefactor3000\relax}%
\providecommand \BibitemShut  [1]{\csname bibitem#1\endcsname}%
\let\auto@bib@innerbib\@empty
\bibitem [{\citenamefont {ik~Bouwmeester}\ \emph {et~al.}(-
  1997/12/11/online)\citenamefont {ik~Bouwmeester}, \citenamefont {ian-W.~ei
  Pan}, \citenamefont {laus Mattle}, \citenamefont {anfred Eibl}, \citenamefont
  {arald Weinfurter},\ and\ \citenamefont {nton
  Zeilinger}}]{1997/12/11/online}%
  \BibitemOpen
  \bibfield  {author} {\bibinfo {author} {\bibfnamefont {D.}~\bibnamefont
  {ik~Bouwmeester}}, \bibinfo {author} {\bibfnamefont {J.}~\bibnamefont
  {ian-W.~ei Pan}}, \bibinfo {author} {\bibfnamefont {K.}~\bibnamefont {laus
  Mattle}}, \bibinfo {author} {\bibfnamefont {M.}~\bibnamefont {anfred Eibl}},
  \bibinfo {author} {\bibfnamefont {H.}~\bibnamefont {arald Weinfurter}}, \
  and\ \bibinfo {author} {\bibfnamefont {A.}~\bibnamefont {nton Zeilinger}},\
  }\bibfield  {title} {\enquote {\bibinfo {title} {- {Experimental} quantum
  teleportation},}\ }\href@noop {} {\ \textbf {\bibinfo {volume} {- 390}},\
  \bibinfo {pages} {-- --} (\bibinfo {year} {- 1997/12/11/online})}\BibitemShut
  {NoStop}%
\bibitem [{\citenamefont {Jennewein}\ \emph {et~al.}(2001)\citenamefont
  {Jennewein}, \citenamefont {Weihs}, \citenamefont {Pan},\ and\ \citenamefont
  {Zeilinger}}]{PhysRevLett.88.017903}%
  \BibitemOpen
  \bibfield  {author} {\bibinfo {author} {\bibfnamefont {Thomas}\ \bibnamefont
  {Jennewein}}, \bibinfo {author} {\bibfnamefont {Gregor}\ \bibnamefont
  {Weihs}}, \bibinfo {author} {\bibfnamefont {Jian-Wei}\ \bibnamefont {Pan}}, \
  and\ \bibinfo {author} {\bibfnamefont {Anton}\ \bibnamefont {Zeilinger}},\
  }\bibfield  {title} {\enquote {\bibinfo {title} {Experimental nonlocality
  proof of quantum teleportation and entanglement swapping},}\ }\href {\doibase
  10.1103/PhysRevLett.88.017903} {\bibfield  {journal} {\bibinfo  {journal}
  {Phys. Rev. Lett.}\ }\textbf {\bibinfo {volume} {88}},\ \bibinfo {pages}
  {017903} (\bibinfo {year} {2001})}\BibitemShut {NoStop}%
\bibitem [{\citenamefont {Cerf}\ \emph {et~al.}(2007)\citenamefont {Cerf},
  \citenamefont {Cerf}, \citenamefont {Leuchs},\ and\ \citenamefont
  {Polzik}}]{cerf2007quantum}%
  \BibitemOpen
  \bibfield  {author} {\bibinfo {author} {\bibfnamefont {N.J.}\ \bibnamefont
  {Cerf}}, \bibinfo {author} {\bibfnamefont {N.J.}\ \bibnamefont {Cerf}},
  \bibinfo {author} {\bibfnamefont {G.}~\bibnamefont {Leuchs}}, \ and\ \bibinfo
  {author} {\bibfnamefont {E.S.}\ \bibnamefont {Polzik}},\ }\href@noop {}
  {\emph {\bibinfo {title} {Quantum Information with Continuous Variables of
  Atoms and Light}}}\ (\bibinfo  {publisher} {Imperial College Press},\
  \bibinfo {year} {2007})\BibitemShut {NoStop}%
\bibitem [{\citenamefont {Bouwmeester}\ \emph {et~al.}(2013)\citenamefont
  {Bouwmeester}, \citenamefont {Ekert},\ and\ \citenamefont
  {Zeilinger}}]{bouwmeester2013physics}%
  \BibitemOpen
  \bibfield  {author} {\bibinfo {author} {\bibfnamefont {D.}~\bibnamefont
  {Bouwmeester}}, \bibinfo {author} {\bibfnamefont {A.K.}\ \bibnamefont
  {Ekert}}, \ and\ \bibinfo {author} {\bibfnamefont {A.}~\bibnamefont
  {Zeilinger}},\ }\href@noop {} {\emph {\bibinfo {title} {The Physics of
  Quantum Information: Quantum Cryptography, Quantum Teleportation, Quantum
  Computation}}}\ (\bibinfo  {publisher} {Springer Berlin Heidelberg},\
  \bibinfo {year} {2013})\BibitemShut {NoStop}%
\bibitem [{\citenamefont {Huang}\ \emph {et~al.}(2016)\citenamefont {Huang},
  \citenamefont {Huang}, \citenamefont {Lin},\ and\ \citenamefont
  {Zeng}}]{Huang2016}%
  \BibitemOpen
  \bibfield  {author} {\bibinfo {author} {\bibfnamefont {Duan}\ \bibnamefont
  {Huang}}, \bibinfo {author} {\bibfnamefont {Peng}\ \bibnamefont {Huang}},
  \bibinfo {author} {\bibfnamefont {Dakai}\ \bibnamefont {Lin}}, \ and\
  \bibinfo {author} {\bibfnamefont {Guihua}\ \bibnamefont {Zeng}},\ }\bibfield
  {title} {\enquote {\bibinfo {title} {Long-distance continuous-variable
  quantum key distribution by controlling excess noise},}\ }\href
  {http://dx.doi.org/10.1038/srep19201} {\bibfield  {journal} {\bibinfo
  {journal} {Scientific Reports}\ }\textbf {\bibinfo {volume} {6}},\ \bibinfo
  {pages} {19201 EP --} (\bibinfo {year} {2016})},\ \bibinfo {note}
  {article}\BibitemShut {NoStop}%
\bibitem [{\citenamefont {Leverrier}\ and\ \citenamefont
  {Grangier}(2011)}]{PhysRevA.83.042312}%
  \BibitemOpen
  \bibfield  {author} {\bibinfo {author} {\bibfnamefont {Anthony}\ \bibnamefont
  {Leverrier}}\ and\ \bibinfo {author} {\bibfnamefont {Philippe}\ \bibnamefont
  {Grangier}},\ }\bibfield  {title} {\enquote {\bibinfo {title}
  {Continuous-variable quantum-key-distribution protocols with a non-gaussian
  modulation},}\ }\href {\doibase 10.1103/PhysRevA.83.042312} {\bibfield
  {journal} {\bibinfo  {journal} {Phys. Rev. A}\ }\textbf {\bibinfo {volume}
  {83}},\ \bibinfo {pages} {042312} (\bibinfo {year} {2011})}\BibitemShut
  {NoStop}%
\bibitem [{\citenamefont {Soh}\ \emph {et~al.}(2015)\citenamefont {Soh},
  \citenamefont {Brif}, \citenamefont {Coles}, \citenamefont {L\"utkenhaus},
  \citenamefont {Camacho}, \citenamefont {Urayama},\ and\ \citenamefont
  {Sarovar}}]{PhysRevX.5.041010}%
  \BibitemOpen
  \bibfield  {author} {\bibinfo {author} {\bibfnamefont {Daniel B.~S.}\
  \bibnamefont {Soh}}, \bibinfo {author} {\bibfnamefont {Constantin}\
  \bibnamefont {Brif}}, \bibinfo {author} {\bibfnamefont {Patrick~J.}\
  \bibnamefont {Coles}}, \bibinfo {author} {\bibfnamefont {Norbert}\
  \bibnamefont {L\"utkenhaus}}, \bibinfo {author} {\bibfnamefont {Ryan~M.}\
  \bibnamefont {Camacho}}, \bibinfo {author} {\bibfnamefont {Junji}\
  \bibnamefont {Urayama}}, \ and\ \bibinfo {author} {\bibfnamefont {Mohan}\
  \bibnamefont {Sarovar}},\ }\bibfield  {title} {\enquote {\bibinfo {title}
  {Self-referenced continuous-variable quantum key distribution protocol},}\
  }\href {\doibase 10.1103/PhysRevX.5.041010} {\bibfield  {journal} {\bibinfo
  {journal} {Phys. Rev. X}\ }\textbf {\bibinfo {volume} {5}},\ \bibinfo {pages}
  {041010} (\bibinfo {year} {2015})}\BibitemShut {NoStop}%
\bibitem [{\citenamefont {Jia}\ \emph {et~al.}(2004)\citenamefont {Jia},
  \citenamefont {Su}, \citenamefont {Pan}, \citenamefont {Gao}, \citenamefont
  {Xie},\ and\ \citenamefont {Peng}}]{PhysRevLett.93.250503}%
  \BibitemOpen
  \bibfield  {author} {\bibinfo {author} {\bibfnamefont {Xiaojun}\ \bibnamefont
  {Jia}}, \bibinfo {author} {\bibfnamefont {Xiaolong}\ \bibnamefont {Su}},
  \bibinfo {author} {\bibfnamefont {Qing}\ \bibnamefont {Pan}}, \bibinfo
  {author} {\bibfnamefont {Jiangrui}\ \bibnamefont {Gao}}, \bibinfo {author}
  {\bibfnamefont {Changde}\ \bibnamefont {Xie}}, \ and\ \bibinfo {author}
  {\bibfnamefont {Kunchi}\ \bibnamefont {Peng}},\ }\bibfield  {title} {\enquote
  {\bibinfo {title} {Experimental demonstration of unconditional entanglement
  swapping for continuous variables},}\ }\href {\doibase
  10.1103/PhysRevLett.93.250503} {\bibfield  {journal} {\bibinfo  {journal}
  {Phys. Rev. Lett.}\ }\textbf {\bibinfo {volume} {93}},\ \bibinfo {pages}
  {250503} (\bibinfo {year} {2004})}\BibitemShut {NoStop}%
\bibitem [{\citenamefont {Miwa}\ \emph {et~al.}(2009)\citenamefont {Miwa},
  \citenamefont {Yoshikawa}, \citenamefont {van Loock},\ and\ \citenamefont
  {Furusawa}}]{PhysRevA.80.050303}%
  \BibitemOpen
  \bibfield  {author} {\bibinfo {author} {\bibfnamefont {Yoshichika}\
  \bibnamefont {Miwa}}, \bibinfo {author} {\bibfnamefont {Jun-ichi}\
  \bibnamefont {Yoshikawa}}, \bibinfo {author} {\bibfnamefont {Peter}\
  \bibnamefont {van Loock}}, \ and\ \bibinfo {author} {\bibfnamefont {Akira}\
  \bibnamefont {Furusawa}},\ }\bibfield  {title} {\enquote {\bibinfo {title}
  {Demonstration of a universal one-way quantum quadratic phase gate},}\ }\href
  {\doibase 10.1103/PhysRevA.80.050303} {\bibfield  {journal} {\bibinfo
  {journal} {Phys. Rev. A}\ }\textbf {\bibinfo {volume} {80}},\ \bibinfo
  {pages} {050303} (\bibinfo {year} {2009})}\BibitemShut {NoStop}%
\bibitem [{\citenamefont {U'Ren}\ \emph {et~al.}(2006)\citenamefont {U'Ren},
  \citenamefont {Erdmann}, \citenamefont {de~la Cruz-Gutierrez},\ and\
  \citenamefont {Walmsley}}]{PhysRevLett.97.223602}%
  \BibitemOpen
  \bibfield  {author} {\bibinfo {author} {\bibfnamefont {Alfred~B.}\
  \bibnamefont {U'Ren}}, \bibinfo {author} {\bibfnamefont {Reinhard~K.}\
  \bibnamefont {Erdmann}}, \bibinfo {author} {\bibfnamefont {Manuel}\
  \bibnamefont {de~la Cruz-Gutierrez}}, \ and\ \bibinfo {author} {\bibfnamefont
  {Ian~A.}\ \bibnamefont {Walmsley}},\ }\bibfield  {title} {\enquote {\bibinfo
  {title} {Generation of two-photon states with an arbitrary degree of
  entanglement via nonlinear crystal superlattices},}\ }\href {\doibase
  10.1103/PhysRevLett.97.223602} {\bibfield  {journal} {\bibinfo  {journal}
  {Phys. Rev. Lett.}\ }\textbf {\bibinfo {volume} {97}},\ \bibinfo {pages}
  {223602} (\bibinfo {year} {2006})}\BibitemShut {NoStop}%
\bibitem [{\citenamefont {Sciscione}\ \emph {et~al.}(2006)\citenamefont
  {Sciscione}, \citenamefont {Centini}, \citenamefont {Sibilia}, \citenamefont
  {Bertolotti},\ and\ \citenamefont {Scalora}}]{PhysRevA.74.013815}%
  \BibitemOpen
  \bibfield  {author} {\bibinfo {author} {\bibfnamefont {L.}~\bibnamefont
  {Sciscione}}, \bibinfo {author} {\bibfnamefont {M.}~\bibnamefont {Centini}},
  \bibinfo {author} {\bibfnamefont {C.}~\bibnamefont {Sibilia}}, \bibinfo
  {author} {\bibfnamefont {M.}~\bibnamefont {Bertolotti}}, \ and\ \bibinfo
  {author} {\bibfnamefont {M.}~\bibnamefont {Scalora}},\ }\bibfield  {title}
  {\enquote {\bibinfo {title} {Entangled, guided photon generation in
  $(1+1)$-dimensional photonic crystals},}\ }\href {\doibase
  10.1103/PhysRevA.74.013815} {\bibfield  {journal} {\bibinfo  {journal} {Phys.
  Rev. A}\ }\textbf {\bibinfo {volume} {74}},\ \bibinfo {pages} {013815}
  (\bibinfo {year} {2006})}\BibitemShut {NoStop}%
\bibitem [{\citenamefont {Yang}\ and\ \citenamefont {Sipe}(2007)}]{Yang:07}%
  \BibitemOpen
  \bibfield  {author} {\bibinfo {author} {\bibfnamefont {Zhenshan}\
  \bibnamefont {Yang}}\ and\ \bibinfo {author} {\bibfnamefont {J.~E.}\
  \bibnamefont {Sipe}},\ }\bibfield  {title} {\enquote {\bibinfo {title}
  {Generating entangled photons via enhanced spontaneous parametric
  downconversion in algaas microring resonators},}\ }\href {\doibase
  10.1364/OL.32.003296} {\bibfield  {journal} {\bibinfo  {journal} {Opt.
  Lett.}\ }\textbf {\bibinfo {volume} {32}},\ \bibinfo {pages} {3296--3298}
  (\bibinfo {year} {2007})}\BibitemShut {NoStop}%
\bibitem [{\citenamefont {Tan}\ \emph {et~al.}(2011)\citenamefont {Tan},
  \citenamefont {Zhang},\ and\ \citenamefont {Li}}]{PhysRevA.83.062310}%
  \BibitemOpen
  \bibfield  {author} {\bibinfo {author} {\bibfnamefont {Hua-Tang}\
  \bibnamefont {Tan}}, \bibinfo {author} {\bibfnamefont {Wei-Min}\ \bibnamefont
  {Zhang}}, \ and\ \bibinfo {author} {\bibfnamefont {Gao-xiang}\ \bibnamefont
  {Li}},\ }\bibfield  {title} {\enquote {\bibinfo {title} {Entangling two
  distant nanocavities via a waveguide},}\ }\href {\doibase
  10.1103/PhysRevA.83.062310} {\bibfield  {journal} {\bibinfo  {journal} {Phys.
  Rev. A}\ }\textbf {\bibinfo {volume} {83}},\ \bibinfo {pages} {062310}
  (\bibinfo {year} {2011})}\BibitemShut {NoStop}%
\bibitem [{\citenamefont {Xiong}\ \emph {et~al.}(2010)\citenamefont {Xiong},
  \citenamefont {Zhang}, \citenamefont {Wang},\ and\ \citenamefont
  {Wu}}]{PhysRevA.82.012105}%
  \BibitemOpen
  \bibfield  {author} {\bibinfo {author} {\bibfnamefont {Heng-Na}\ \bibnamefont
  {Xiong}}, \bibinfo {author} {\bibfnamefont {Wei-Min}\ \bibnamefont {Zhang}},
  \bibinfo {author} {\bibfnamefont {Xiaoguang}\ \bibnamefont {Wang}}, \ and\
  \bibinfo {author} {\bibfnamefont {Meng-Hsiu}\ \bibnamefont {Wu}},\ }\bibfield
   {title} {\enquote {\bibinfo {title} {Exact non-markovian cavity dynamics
  strongly coupled to a reservoir},}\ }\href {\doibase
  10.1103/PhysRevA.82.012105} {\bibfield  {journal} {\bibinfo  {journal} {Phys.
  Rev. A}\ }\textbf {\bibinfo {volume} {82}},\ \bibinfo {pages} {012105}
  (\bibinfo {year} {2010})}\BibitemShut {NoStop}%
\bibitem [{\citenamefont {Tan}\ and\ \citenamefont
  {Zhang}(2011)}]{PhysRevA.83.032102}%
  \BibitemOpen
  \bibfield  {author} {\bibinfo {author} {\bibfnamefont {Hua-Tang}\
  \bibnamefont {Tan}}\ and\ \bibinfo {author} {\bibfnamefont {Wei-Min}\
  \bibnamefont {Zhang}},\ }\bibfield  {title} {\enquote {\bibinfo {title}
  {Non-markovian dynamics of an open quantum system with initial
  system-reservoir correlations: A nanocavity coupled to a coupled-resonator
  optical waveguide},}\ }\href {\doibase 10.1103/PhysRevA.83.032102} {\bibfield
   {journal} {\bibinfo  {journal} {Phys. Rev. A}\ }\textbf {\bibinfo {volume}
  {83}},\ \bibinfo {pages} {032102} (\bibinfo {year} {2011})}\BibitemShut
  {NoStop}%
\bibitem [{\citenamefont {Caillet}\ \emph {et~al.}(2009)\citenamefont
  {Caillet}, \citenamefont {Berger}, \citenamefont {Leo},\ and\ \citenamefont
  {Ducci}}]{doi:10.1080/09500340802192431}%
  \BibitemOpen
  \bibfield  {author} {\bibinfo {author} {\bibfnamefont {X.}~\bibnamefont
  {Caillet}}, \bibinfo {author} {\bibfnamefont {V.}~\bibnamefont {Berger}},
  \bibinfo {author} {\bibfnamefont {G.}~\bibnamefont {Leo}}, \ and\ \bibinfo
  {author} {\bibfnamefont {S.}~\bibnamefont {Ducci}},\ }\bibfield  {title}
  {\enquote {\bibinfo {title} {A semiconductor source of counterpropagating
  twin photons: a versatile device allowing the control of the two-photon
  state},}\ }\href {\doibase 10.1080/09500340802192431} {\bibfield  {journal}
  {\bibinfo  {journal} {Journal of Modern Optics}\ }\textbf {\bibinfo {volume}
  {56}},\ \bibinfo {pages} {232--239} (\bibinfo {year} {2009})},\ \Eprint
  {http://arxiv.org/abs/https://doi.org/10.1080/09500340802192431}
  {https://doi.org/10.1080/09500340802192431} \BibitemShut {NoStop}%
\bibitem [{\citenamefont {Orieux}\ \emph {et~al.}(2011)\citenamefont {Orieux},
  \citenamefont {Caillet}, \citenamefont {Lema\^{i}tre}, \citenamefont
  {Filloux}, \citenamefont {Favero}, \citenamefont {Leo},\ and\ \citenamefont
  {Ducci}}]{Orieux:11}%
  \BibitemOpen
  \bibfield  {author} {\bibinfo {author} {\bibfnamefont {Adeline}\ \bibnamefont
  {Orieux}}, \bibinfo {author} {\bibfnamefont {Xavier}\ \bibnamefont
  {Caillet}}, \bibinfo {author} {\bibfnamefont {Aristide}\ \bibnamefont
  {Lema\^{i}tre}}, \bibinfo {author} {\bibfnamefont {Pascal}\ \bibnamefont
  {Filloux}}, \bibinfo {author} {\bibfnamefont {Ivan}\ \bibnamefont {Favero}},
  \bibinfo {author} {\bibfnamefont {Giuseppe}\ \bibnamefont {Leo}}, \ and\
  \bibinfo {author} {\bibfnamefont {Sara}\ \bibnamefont {Ducci}},\ }\bibfield
  {title} {\enquote {\bibinfo {title} {Efficient parametric generation of
  counterpropagating two-photon states},}\ }\href {\doibase
  10.1364/JOSAB.28.000045} {\bibfield  {journal} {\bibinfo  {journal} {J. Opt.
  Soc. Am. B}\ }\textbf {\bibinfo {volume} {28}},\ \bibinfo {pages} {45--51}
  (\bibinfo {year} {2011})}\BibitemShut {NoStop}%
\bibitem [{\citenamefont {Yariv}\ \emph {et~al.}(1999)\citenamefont {Yariv},
  \citenamefont {Xu}, \citenamefont {Lee},\ and\ \citenamefont
  {Scherer}}]{Yariv:99}%
  \BibitemOpen
  \bibfield  {author} {\bibinfo {author} {\bibfnamefont {Amnon}\ \bibnamefont
  {Yariv}}, \bibinfo {author} {\bibfnamefont {Yong}\ \bibnamefont {Xu}},
  \bibinfo {author} {\bibfnamefont {Reginald~K.}\ \bibnamefont {Lee}}, \ and\
  \bibinfo {author} {\bibfnamefont {Axel}\ \bibnamefont {Scherer}},\ }\bibfield
   {title} {\enquote {\bibinfo {title} {Coupled-resonator optical waveguide:?a
  proposal and analysis},}\ }\href {\doibase 10.1364/OL.24.000711} {\bibfield
  {journal} {\bibinfo  {journal} {Opt. Lett.}\ }\textbf {\bibinfo {volume}
  {24}},\ \bibinfo {pages} {711--713} (\bibinfo {year} {1999})}\BibitemShut
  {NoStop}%
\bibitem [{\citenamefont {Slusher}\ \emph {et~al.}(1985)\citenamefont
  {Slusher}, \citenamefont {Hollberg}, \citenamefont {Yurke}, \citenamefont
  {Mertz},\ and\ \citenamefont {Valley}}]{PhysRevLett.55.2409}%
  \BibitemOpen
  \bibfield  {author} {\bibinfo {author} {\bibfnamefont {R.~E.}\ \bibnamefont
  {Slusher}}, \bibinfo {author} {\bibfnamefont {L.~W.}\ \bibnamefont
  {Hollberg}}, \bibinfo {author} {\bibfnamefont {B.}~\bibnamefont {Yurke}},
  \bibinfo {author} {\bibfnamefont {J.~C.}\ \bibnamefont {Mertz}}, \ and\
  \bibinfo {author} {\bibfnamefont {J.~F.}\ \bibnamefont {Valley}},\ }\bibfield
   {title} {\enquote {\bibinfo {title} {Observation of squeezed states
  generated by four-wave mixing in an optical cavity},}\ }\href {\doibase
  10.1103/PhysRevLett.55.2409} {\bibfield  {journal} {\bibinfo  {journal}
  {Phys. Rev. Lett.}\ }\textbf {\bibinfo {volume} {55}},\ \bibinfo {pages}
  {2409--2412} (\bibinfo {year} {1985})}\BibitemShut {NoStop}%
\bibitem [{\citenamefont {Jasperse}\ \emph {et~al.}(2011)\citenamefont
  {Jasperse}, \citenamefont {Turner},\ and\ \citenamefont
  {Scholten}}]{Jasperse:11}%
  \BibitemOpen
  \bibfield  {author} {\bibinfo {author} {\bibfnamefont {M.}~\bibnamefont
  {Jasperse}}, \bibinfo {author} {\bibfnamefont {L.~D.}\ \bibnamefont
  {Turner}}, \ and\ \bibinfo {author} {\bibfnamefont {R.~E.}\ \bibnamefont
  {Scholten}},\ }\bibfield  {title} {\enquote {\bibinfo {title} {Relative
  intensity squeezing by four-wave mixing with loss: an analytic model and
  experimental diagnostic},}\ }\href {\doibase 10.1364/OE.19.003765} {\bibfield
   {journal} {\bibinfo  {journal} {Opt. Express}\ }\textbf {\bibinfo {volume}
  {19}},\ \bibinfo {pages} {3765--3774} (\bibinfo {year} {2011})}\BibitemShut
  {NoStop}%
\bibitem [{\citenamefont {Seifoory}\ \emph {et~al.}(2017)\citenamefont
  {Seifoory}, \citenamefont {Doutre}, \citenamefont {Dignam},\ and\
  \citenamefont {Sipe}}]{Seifoory:17}%
  \BibitemOpen
  \bibfield  {author} {\bibinfo {author} {\bibfnamefont {Hossein}\ \bibnamefont
  {Seifoory}}, \bibinfo {author} {\bibfnamefont {Sean}\ \bibnamefont {Doutre}},
  \bibinfo {author} {\bibfnamefont {Marc.~M.}\ \bibnamefont {Dignam}}, \ and\
  \bibinfo {author} {\bibfnamefont {J.~E.}\ \bibnamefont {Sipe}},\ }\bibfield
  {title} {\enquote {\bibinfo {title} {Squeezed thermal states: the result of
  parametric down conversion in lossy cavities},}\ }\href {\doibase
  10.1364/JOSAB.34.001587} {\bibfield  {journal} {\bibinfo  {journal} {J. Opt.
  Soc. Am. B}\ }\textbf {\bibinfo {volume} {34}},\ \bibinfo {pages}
  {1587--1596} (\bibinfo {year} {2017})}\BibitemShut {NoStop}%
\bibitem [{\citenamefont {Seifoory}\ and\ \citenamefont
  {Dignam}(2018)}]{PhysRevA.97.023840}%
  \BibitemOpen
  \bibfield  {author} {\bibinfo {author} {\bibfnamefont {Hossein}\ \bibnamefont
  {Seifoory}}\ and\ \bibinfo {author} {\bibfnamefont {Marc~M.}\ \bibnamefont
  {Dignam}},\ }\bibfield  {title} {\enquote {\bibinfo {title} {Squeezed-state
  evolution and entanglement in lossy coupled-resonator optical waveguides},}\
  }\href {\doibase 10.1103/PhysRevA.97.023840} {\bibfield  {journal} {\bibinfo
  {journal} {Phys. Rev. A}\ }\textbf {\bibinfo {volume} {97}},\ \bibinfo
  {pages} {023840} (\bibinfo {year} {2018})}\BibitemShut {NoStop}%
\bibitem [{\citenamefont {Rai}\ and\ \citenamefont
  {Angelakis}(2012)}]{PhysRevA.85.052330}%
  \BibitemOpen
  \bibfield  {author} {\bibinfo {author} {\bibfnamefont {Amit}\ \bibnamefont
  {Rai}}\ and\ \bibinfo {author} {\bibfnamefont {Dimitris~G.}\ \bibnamefont
  {Angelakis}},\ }\bibfield  {title} {\enquote {\bibinfo {title} {Dynamics of
  nonclassical light in integrated nonlinear waveguide arrays and generation of
  robust continuous-variable entanglement},}\ }\href {\doibase
  10.1103/PhysRevA.85.052330} {\bibfield  {journal} {\bibinfo  {journal} {Phys.
  Rev. A}\ }\textbf {\bibinfo {volume} {85}},\ \bibinfo {pages} {052330}
  (\bibinfo {year} {2012})}\BibitemShut {NoStop}%
\bibitem [{\citenamefont {Walton}\ \emph {et~al.}(2003)\citenamefont {Walton},
  \citenamefont {Booth}, \citenamefont {Sergienko}, \citenamefont {Saleh},\
  and\ \citenamefont {Teich}}]{PhysRevA.67.053810}%
  \BibitemOpen
  \bibfield  {author} {\bibinfo {author} {\bibfnamefont {Zachary~D.}\
  \bibnamefont {Walton}}, \bibinfo {author} {\bibfnamefont {Mark~C.}\
  \bibnamefont {Booth}}, \bibinfo {author} {\bibfnamefont {Alexander~V.}\
  \bibnamefont {Sergienko}}, \bibinfo {author} {\bibfnamefont {Bahaa E.~A.}\
  \bibnamefont {Saleh}}, \ and\ \bibinfo {author} {\bibfnamefont {Malvin~C.}\
  \bibnamefont {Teich}},\ }\bibfield  {title} {\enquote {\bibinfo {title}
  {Controllable frequency entanglement via auto-phase-matched spontaneous
  parametric down-conversion},}\ }\href {\doibase 10.1103/PhysRevA.67.053810}
  {\bibfield  {journal} {\bibinfo  {journal} {Phys. Rev. A}\ }\textbf {\bibinfo
  {volume} {67}},\ \bibinfo {pages} {053810} (\bibinfo {year}
  {2003})}\BibitemShut {NoStop}%
\bibitem [{\citenamefont {Walton}\ \emph {et~al.}(2004)\citenamefont {Walton},
  \citenamefont {Sergienko}, \citenamefont {Saleh},\ and\ \citenamefont
  {Teich}}]{PhysRevA.70.052317}%
  \BibitemOpen
  \bibfield  {author} {\bibinfo {author} {\bibfnamefont {Zachary~D.}\
  \bibnamefont {Walton}}, \bibinfo {author} {\bibfnamefont {Alexander~V.}\
  \bibnamefont {Sergienko}}, \bibinfo {author} {\bibfnamefont {Bahaa E.~A.}\
  \bibnamefont {Saleh}}, \ and\ \bibinfo {author} {\bibfnamefont {Malvin~C.}\
  \bibnamefont {Teich}},\ }\bibfield  {title} {\enquote {\bibinfo {title}
  {Generation of polarization-entangled photon pairs with arbitrary joint
  spectrum},}\ }\href {\doibase 10.1103/PhysRevA.70.052317} {\bibfield
  {journal} {\bibinfo  {journal} {Phys. Rev. A}\ }\textbf {\bibinfo {volume}
  {70}},\ \bibinfo {pages} {052317} (\bibinfo {year} {2004})}\BibitemShut
  {NoStop}%
\bibitem [{\citenamefont {Ashcroft}\ and\ \citenamefont
  {Mermin}(2011)}]{ashcroft2011solid}%
  \BibitemOpen
  \bibfield  {author} {\bibinfo {author} {\bibfnamefont {N.W.}\ \bibnamefont
  {Ashcroft}}\ and\ \bibinfo {author} {\bibfnamefont {N.D.}\ \bibnamefont
  {Mermin}},\ }\href {https://books.google.ca/books?id=x\_s\_YAAACAAJ} {\emph
  {\bibinfo {title} {Solid State Physics}}}\ (\bibinfo  {publisher} {Cengage
  Learning},\ \bibinfo {year} {2011})\BibitemShut {NoStop}%
\bibitem [{\citenamefont {Fussell}\ and\ \citenamefont
  {Dignam}(2007{\natexlab{a}})}]{doi:10.1063/1.2737430}%
  \BibitemOpen
  \bibfield  {author} {\bibinfo {author} {\bibfnamefont {David~P.}\
  \bibnamefont {Fussell}}\ and\ \bibinfo {author} {\bibfnamefont {Marc~M.}\
  \bibnamefont {Dignam}},\ }\bibfield  {title} {\enquote {\bibinfo {title}
  {Engineering the quality factors of coupled-cavity modes in photonic crystal
  slabs},}\ }\href {\doibase 10.1063/1.2737430} {\bibfield  {journal} {\bibinfo
   {journal} {Applied Physics Letters}\ }\textbf {\bibinfo {volume} {90}},\
  \bibinfo {pages} {183121} (\bibinfo {year} {2007}{\natexlab{a}})},\ \Eprint
  {http://arxiv.org/abs/http://dx.doi.org/10.1063/1.2737430}
  {http://dx.doi.org/10.1063/1.2737430} \BibitemShut {NoStop}%
\bibitem [{com()}]{complex}%
  \BibitemOpen
  \href@noop {} {}\bibinfo {note} {Throughout this work, we place a tilde on
  top of quantities that are inherently complex (as opposed to
  real).}\BibitemShut {Stop}%
\bibitem [{\citenamefont {Kamandar~Dezfouli}(2015)}]{MohsenThesis}%
  \BibitemOpen
  \bibfield  {author} {\bibinfo {author} {\bibfnamefont {Mohsen}\ \bibnamefont
  {Kamandar~Dezfouli}},\ }\emph {\bibinfo {title} {Quantum Nonlinear Optics in
  Lossy Coupled-Cavities in Photonic Crystal Slabs}},\ \href@noop {} {Ph.D.
  thesis},\ \bibinfo  {school} {Queen's University} (\bibinfo {year}
  {2015})\BibitemShut {NoStop}%
\bibitem [{\citenamefont {Fussell}\ and\ \citenamefont
  {Dignam}(2007{\natexlab{b}})}]{Fussell:07}%
  \BibitemOpen
  \bibfield  {author} {\bibinfo {author} {\bibfnamefont {David~P.}\
  \bibnamefont {Fussell}}\ and\ \bibinfo {author} {\bibfnamefont {Marc~M.}\
  \bibnamefont {Dignam}},\ }\bibfield  {title} {\enquote {\bibinfo {title}
  {Spontaneous emission in coupled microcavity-waveguide structures at the band
  edge},}\ }\href {\doibase 10.1364/OL.32.001527} {\bibfield  {journal}
  {\bibinfo  {journal} {Opt. Lett.}\ }\textbf {\bibinfo {volume} {32}},\
  \bibinfo {pages} {1527--1529} (\bibinfo {year}
  {2007}{\natexlab{b}})}\BibitemShut {NoStop}%
\bibitem [{\citenamefont {Saravi}\ \emph {et~al.}(2017)\citenamefont {Saravi},
  \citenamefont {Pertsch},\ and\ \citenamefont
  {Setzpfandt}}]{PhysRevLett.118.183603}%
  \BibitemOpen
  \bibfield  {author} {\bibinfo {author} {\bibfnamefont {Sina}\ \bibnamefont
  {Saravi}}, \bibinfo {author} {\bibfnamefont {Thomas}\ \bibnamefont
  {Pertsch}}, \ and\ \bibinfo {author} {\bibfnamefont {Frank}\ \bibnamefont
  {Setzpfandt}},\ }\bibfield  {title} {\enquote {\bibinfo {title} {Generation
  of counterpropagating path-entangled photon pairs in a single periodic
  waveguide},}\ }\href {\doibase 10.1103/PhysRevLett.118.183603} {\bibfield
  {journal} {\bibinfo  {journal} {Phys. Rev. Lett.}\ }\textbf {\bibinfo
  {volume} {118}},\ \bibinfo {pages} {183603} (\bibinfo {year}
  {2017})}\BibitemShut {NoStop}%
\bibitem [{\citenamefont {Yang}\ \emph {et~al.}(2008)\citenamefont {Yang},
  \citenamefont {Liscidini},\ and\ \citenamefont {Sipe}}]{PhysRevA.77.033808}%
  \BibitemOpen
  \bibfield  {author} {\bibinfo {author} {\bibfnamefont {Zhenshan}\
  \bibnamefont {Yang}}, \bibinfo {author} {\bibfnamefont {Marco}\ \bibnamefont
  {Liscidini}}, \ and\ \bibinfo {author} {\bibfnamefont {J.~E.}\ \bibnamefont
  {Sipe}},\ }\bibfield  {title} {\enquote {\bibinfo {title} {Spontaneous
  parametric down-conversion in waveguides: A backward heisenberg picture
  approach},}\ }\href {\doibase 10.1103/PhysRevA.77.033808} {\bibfield
  {journal} {\bibinfo  {journal} {Phys. Rev. A}\ }\textbf {\bibinfo {volume}
  {77}},\ \bibinfo {pages} {033808} (\bibinfo {year} {2008})}\BibitemShut
  {NoStop}%
\bibitem [{\citenamefont {Peres}(2006)}]{peres2006quantum}%
  \BibitemOpen
  \bibfield  {author} {\bibinfo {author} {\bibfnamefont {A.}~\bibnamefont
  {Peres}},\ }\href {https://books.google.ca/books?id=pQXSBwAAQBAJ} {\emph
  {\bibinfo {title} {Quantum Theory: Concepts and Methods}}},\ Fundamental
  Theories of Physics\ (\bibinfo  {publisher} {Springer Netherlands},\ \bibinfo
  {year} {2006})\BibitemShut {NoStop}%
\bibitem [{\citenamefont {Ekert}\ and\ \citenamefont
  {Knight}(1995)}]{doi:10.1119/1.17904}%
  \BibitemOpen
  \bibfield  {author} {\bibinfo {author} {\bibfnamefont {Artur}\ \bibnamefont
  {Ekert}}\ and\ \bibinfo {author} {\bibfnamefont {Peter~L.}\ \bibnamefont
  {Knight}},\ }\bibfield  {title} {\enquote {\bibinfo {title} {Entangled
  quantum systems and the schmidt decomposition},}\ }\href {\doibase
  10.1119/1.17904} {\bibfield  {journal} {\bibinfo  {journal} {American Journal
  of Physics}\ }\textbf {\bibinfo {volume} {63}},\ \bibinfo {pages} {415--423}
  (\bibinfo {year} {1995})},\ \Eprint
  {http://arxiv.org/abs/https://doi.org/10.1119/1.17904}
  {https://doi.org/10.1119/1.17904} \BibitemShut {NoStop}%
\bibitem [{\citenamefont {Breuer}\ and\ \citenamefont
  {Petruccione}(2007)}]{open_quantum_system}%
  \BibitemOpen
  \bibfield  {author} {\bibinfo {author} {\bibfnamefont {H.~P.}\ \bibnamefont
  {Breuer}}\ and\ \bibinfo {author} {\bibfnamefont {F.}~\bibnamefont
  {Petruccione}},\ }\href@noop {} {\emph {\bibinfo {title} {The theory of open
  quantum systems}}}\ (\bibinfo  {publisher} {Oxford University Press},\
  \bibinfo {year} {2007})\BibitemShut {NoStop}%
\bibitem [{\citenamefont {Dignam}\ and\ \citenamefont
  {Dezfouli}(2012)}]{PhysRevA.85.013809}%
  \BibitemOpen
  \bibfield  {author} {\bibinfo {author} {\bibfnamefont {Marc.~M.}\
  \bibnamefont {Dignam}}\ and\ \bibinfo {author} {\bibfnamefont
  {Mohsen~Kamandar}\ \bibnamefont {Dezfouli}},\ }\bibfield  {title} {\enquote
  {\bibinfo {title} {Photon--quantum-dot dynamics in coupled-cavity photonic
  crystal slabs},}\ }\href {\doibase 10.1103/PhysRevA.85.013809} {\bibfield
  {journal} {\bibinfo  {journal} {Phys. Rev. A}\ }\textbf {\bibinfo {volume}
  {85}},\ \bibinfo {pages} {013809} (\bibinfo {year} {2012})}\BibitemShut
  {NoStop}%
\bibitem [{\citenamefont {Duan}\ \emph {et~al.}(2000)\citenamefont {Duan},
  \citenamefont {Giedke}, \citenamefont {Cirac},\ and\ \citenamefont
  {Zoller}}]{PhysRevLett.84.2722}%
  \BibitemOpen
  \bibfield  {author} {\bibinfo {author} {\bibfnamefont {Lu-Ming}\ \bibnamefont
  {Duan}}, \bibinfo {author} {\bibfnamefont {G.}~\bibnamefont {Giedke}},
  \bibinfo {author} {\bibfnamefont {J.~I.}\ \bibnamefont {Cirac}}, \ and\
  \bibinfo {author} {\bibfnamefont {P.}~\bibnamefont {Zoller}},\ }\bibfield
  {title} {\enquote {\bibinfo {title} {Inseparability criterion for continuous
  variable systems},}\ }\href {\doibase 10.1103/PhysRevLett.84.2722} {\bibfield
   {journal} {\bibinfo  {journal} {Phys. Rev. Lett.}\ }\textbf {\bibinfo
  {volume} {84}},\ \bibinfo {pages} {2722--2725} (\bibinfo {year}
  {2000})}\BibitemShut {NoStop}%
\bibitem [{\citenamefont {Simon}(2000)}]{PhysRevLett.84.2726}%
  \BibitemOpen
  \bibfield  {author} {\bibinfo {author} {\bibfnamefont {R.}~\bibnamefont
  {Simon}},\ }\bibfield  {title} {\enquote {\bibinfo {title} {Peres-horodecki
  separability criterion for continuous variable systems},}\ }\href {\doibase
  10.1103/PhysRevLett.84.2726} {\bibfield  {journal} {\bibinfo  {journal}
  {Phys. Rev. Lett.}\ }\textbf {\bibinfo {volume} {84}},\ \bibinfo {pages}
  {2726--2729} (\bibinfo {year} {2000})}\BibitemShut {NoStop}%
\bibitem [{\citenamefont {Masada}\ \emph {et~al.}(2015)\citenamefont {Masada},
  \citenamefont {Miyata}, \citenamefont {Politi}, \citenamefont {Hashimoto},
  \citenamefont {O'Brien},\ and\ \citenamefont {Furusawa}}]{Masada2015}%
  \BibitemOpen
  \bibfield  {author} {\bibinfo {author} {\bibfnamefont {Genta}\ \bibnamefont
  {Masada}}, \bibinfo {author} {\bibfnamefont {Kazunori}\ \bibnamefont
  {Miyata}}, \bibinfo {author} {\bibfnamefont {Alberto}\ \bibnamefont
  {Politi}}, \bibinfo {author} {\bibfnamefont {Toshikazu}\ \bibnamefont
  {Hashimoto}}, \bibinfo {author} {\bibfnamefont {Jeremy~L.}\ \bibnamefont
  {O'Brien}}, \ and\ \bibinfo {author} {\bibfnamefont {Akira}\ \bibnamefont
  {Furusawa}},\ }\bibfield  {title} {\enquote {\bibinfo {title}
  {Continuous-variable entanglement on a chip},}\ }\href
  {http://dx.doi.org/10.1038/nphoton.2015.42} {\bibfield  {journal} {\bibinfo
  {journal} {Nat Photon}\ }\textbf {\bibinfo {volume} {9}},\ \bibinfo {pages}
  {316--319} (\bibinfo {year} {2015})},\ \bibinfo {note} {letter}\BibitemShut
  {NoStop}%
\bibitem [{\citenamefont {Zhang}\ \emph {et~al.}(2015)\citenamefont {Zhang},
  \citenamefont {Okubo}, \citenamefont {Hirano}, \citenamefont {Eto},\ and\
  \citenamefont {Hirano}}]{Zhang2015}%
  \BibitemOpen
  \bibfield  {author} {\bibinfo {author} {\bibfnamefont {Yun}\ \bibnamefont
  {Zhang}}, \bibinfo {author} {\bibfnamefont {Ryuhi}\ \bibnamefont {Okubo}},
  \bibinfo {author} {\bibfnamefont {Mayumi}\ \bibnamefont {Hirano}}, \bibinfo
  {author} {\bibfnamefont {Yujiro}\ \bibnamefont {Eto}}, \ and\ \bibinfo
  {author} {\bibfnamefont {Takuya}\ \bibnamefont {Hirano}},\ }\bibfield
  {title} {\enquote {\bibinfo {title} {Experimental realization of spatially
  separated entanglement with continuous variables using laser pulse trains},}\
  }\href {http://dx.doi.org/10.1038/srep13029} {\bibfield  {journal} {\bibinfo
  {journal} {Scientific Reports}\ }\textbf {\bibinfo {volume} {5}},\ \bibinfo
  {pages} {13029 EP --} (\bibinfo {year} {2015})},\ \bibinfo {note}
  {article}\BibitemShut {NoStop}%
\bibitem [{wel()}]{well_localized}%
  \BibitemOpen
  \href@noop {} {}\bibinfo {note} {By well-localized we mean that the integral
  would be negligible unless both of these modes are at the same
  $p$}\BibitemShut {NoStop}%
\bibitem [{exp()}]{expiq_x}%
  \BibitemOpen
  \href@noop {} {}\bibinfo {note} {For the range of pump frequency considered
  in this paper, to a very good approximation one can consider $e^{iq_xx}\simeq
  e^{iq_px}$.}\BibitemShut {Stop}%
\bibitem [{\citenamefont {U'Ren}\ \emph {et~al.}(2003)\citenamefont {U'Ren},
  \citenamefont {Banaszek},\ and\ \citenamefont
  {Walmsley}}]{U'Ren:2003:PEQ:2011564.2011567}%
  \BibitemOpen
  \bibfield  {author} {\bibinfo {author} {\bibfnamefont {A.~B.}\ \bibnamefont
  {U'Ren}}, \bibinfo {author} {\bibfnamefont {K.}~\bibnamefont {Banaszek}}, \
  and\ \bibinfo {author} {\bibfnamefont {I.~A.}\ \bibnamefont {Walmsley}},\
  }\bibfield  {title} {\enquote {\bibinfo {title} {Photon engineering for
  quantum information processing},}\ }\href
  {http://dl.acm.org/citation.cfm?id=2011564.2011567} {\bibfield  {journal}
  {\bibinfo  {journal} {Quantum Info. Comput.}\ }\textbf {\bibinfo {volume}
  {3}},\ \bibinfo {pages} {480--502} (\bibinfo {year} {2003})}\BibitemShut
  {NoStop}%
\bibitem [{\citenamefont {Lvovsky}\ \emph {et~al.}(2007)\citenamefont
  {Lvovsky}, \citenamefont {Wasilewski},\ and\ \citenamefont
  {Banaszek}}]{doi:10.1080/09500340600777805}%
  \BibitemOpen
  \bibfield  {author} {\bibinfo {author} {\bibfnamefont {A.~I.}\ \bibnamefont
  {Lvovsky}}, \bibinfo {author} {\bibfnamefont {Wojciech}\ \bibnamefont
  {Wasilewski}}, \ and\ \bibinfo {author} {\bibfnamefont {Konrad}\ \bibnamefont
  {Banaszek}},\ }\bibfield  {title} {\enquote {\bibinfo {title} {Decomposing a
  pulsed optical parametric amplifier into independent squeezers},}\ }\href
  {\doibase 10.1080/09500340600777805} {\bibfield  {journal} {\bibinfo
  {journal} {Journal of Modern Optics}\ }\textbf {\bibinfo {volume} {54}},\
  \bibinfo {pages} {721--733} (\bibinfo {year} {2007})},\ \Eprint
  {http://arxiv.org/abs/https://doi.org/10.1080/09500340600777805}
  {https://doi.org/10.1080/09500340600777805} \BibitemShut {NoStop}%
\bibitem [{\citenamefont {Blay}\ \emph {et~al.}(2017)\citenamefont {Blay},
  \citenamefont {Steel},\ and\ \citenamefont {Helt}}]{PhysRevA.96.053842}%
  \BibitemOpen
  \bibfield  {author} {\bibinfo {author} {\bibfnamefont {Daniel~R.}\
  \bibnamefont {Blay}}, \bibinfo {author} {\bibfnamefont {M.~J.}\ \bibnamefont
  {Steel}}, \ and\ \bibinfo {author} {\bibfnamefont {L.~G.}\ \bibnamefont
  {Helt}},\ }\bibfield  {title} {\enquote {\bibinfo {title} {Effects of
  filtering on the purity of heralded single photons from parametric
  sources},}\ }\href {\doibase 10.1103/PhysRevA.96.053842} {\bibfield
  {journal} {\bibinfo  {journal} {Phys. Rev. A}\ }\textbf {\bibinfo {volume}
  {96}},\ \bibinfo {pages} {053842} (\bibinfo {year} {2017})}\BibitemShut
  {NoStop}%
\bibitem [{\citenamefont {Zhukovsky}\ \emph {et~al.}(2012)\citenamefont
  {Zhukovsky}, \citenamefont {Helt}, \citenamefont {Abolghasem}, \citenamefont
  {Kang}, \citenamefont {Sipe},\ and\ \citenamefont {Helmy}}]{Zhukovsky:12}%
  \BibitemOpen
  \bibfield  {author} {\bibinfo {author} {\bibfnamefont {Sergei~V.}\
  \bibnamefont {Zhukovsky}}, \bibinfo {author} {\bibfnamefont {Lukas~G.}\
  \bibnamefont {Helt}}, \bibinfo {author} {\bibfnamefont {Payam}\ \bibnamefont
  {Abolghasem}}, \bibinfo {author} {\bibfnamefont {Dongpeng}\ \bibnamefont
  {Kang}}, \bibinfo {author} {\bibfnamefont {John~E.}\ \bibnamefont {Sipe}}, \
  and\ \bibinfo {author} {\bibfnamefont {Amr~S.}\ \bibnamefont {Helmy}},\
  }\bibfield  {title} {\enquote {\bibinfo {title} {Bragg reflection waveguides
  as integrated sources of entangled photon pairs},}\ }\href {\doibase
  10.1364/JOSAB.29.002516} {\bibfield  {journal} {\bibinfo  {journal} {J. Opt.
  Soc. Am. B}\ }\textbf {\bibinfo {volume} {29}},\ \bibinfo {pages}
  {2516--2523} (\bibinfo {year} {2012})}\BibitemShut {NoStop}%
\bibitem [{\citenamefont {Gili}\ \emph {et~al.}(2016)\citenamefont {Gili},
  \citenamefont {Carletti}, \citenamefont {Locatelli}, \citenamefont {Rocco},
  \citenamefont {Finazzi}, \citenamefont {Ghirardini}, \citenamefont {Favero},
  \citenamefont {Gomez}, \citenamefont {Lema\^{i}tre}, \citenamefont
  {Celebrano}, \citenamefont {Angelis},\ and\ \citenamefont {Leo}}]{Gili:16}%
  \BibitemOpen
  \bibfield  {author} {\bibinfo {author} {\bibfnamefont {V.~F.}\ \bibnamefont
  {Gili}}, \bibinfo {author} {\bibfnamefont {L.}~\bibnamefont {Carletti}},
  \bibinfo {author} {\bibfnamefont {A.}~\bibnamefont {Locatelli}}, \bibinfo
  {author} {\bibfnamefont {D.}~\bibnamefont {Rocco}}, \bibinfo {author}
  {\bibfnamefont {M.}~\bibnamefont {Finazzi}}, \bibinfo {author} {\bibfnamefont
  {L.}~\bibnamefont {Ghirardini}}, \bibinfo {author} {\bibfnamefont
  {I.}~\bibnamefont {Favero}}, \bibinfo {author} {\bibfnamefont
  {C.}~\bibnamefont {Gomez}}, \bibinfo {author} {\bibfnamefont
  {A.}~\bibnamefont {Lema\^{i}tre}}, \bibinfo {author} {\bibfnamefont
  {M.}~\bibnamefont {Celebrano}}, \bibinfo {author} {\bibfnamefont {C.~De}\
  \bibnamefont {Angelis}}, \ and\ \bibinfo {author} {\bibfnamefont
  {G.}~\bibnamefont {Leo}},\ }\bibfield  {title} {\enquote {\bibinfo {title}
  {Monolithic algaas second-harmonic nanoantennas},}\ }\href {\doibase
  10.1364/OE.24.015965} {\bibfield  {journal} {\bibinfo  {journal} {Opt.
  Express}\ }\textbf {\bibinfo {volume} {24}},\ \bibinfo {pages} {15965--15971}
  (\bibinfo {year} {2016})}\BibitemShut {NoStop}%
\bibitem [{\citenamefont {Carletti}\ \emph {et~al.}(2015)\citenamefont
  {Carletti}, \citenamefont {Locatelli}, \citenamefont {Stepanenko},
  \citenamefont {Leo},\ and\ \citenamefont {Angelis}}]{Carletti:15}%
  \BibitemOpen
  \bibfield  {author} {\bibinfo {author} {\bibfnamefont {L.}~\bibnamefont
  {Carletti}}, \bibinfo {author} {\bibfnamefont {A.}~\bibnamefont {Locatelli}},
  \bibinfo {author} {\bibfnamefont {O.}~\bibnamefont {Stepanenko}}, \bibinfo
  {author} {\bibfnamefont {G.}~\bibnamefont {Leo}}, \ and\ \bibinfo {author}
  {\bibfnamefont {C.~De}\ \bibnamefont {Angelis}},\ }\bibfield  {title}
  {\enquote {\bibinfo {title} {Enhanced second-harmonic generation from
  magnetic resonance in algaas nanoantennas},}\ }\href {\doibase
  10.1364/OE.23.026544} {\bibfield  {journal} {\bibinfo  {journal} {Opt.
  Express}\ }\textbf {\bibinfo {volume} {23}},\ \bibinfo {pages} {26544--26550}
  (\bibinfo {year} {2015})}\BibitemShut {NoStop}%
\end{thebibliography}%
\end{document}